\documentclass[12pt]{iopart}

\usepackage{ifthen}
\usepackage{ifpdf}
\usepackage{color}

\usepackage[section]{placeins}

\ifpdf
\usepackage{graphicx}
\usepackage{epstopdf}
\else
\usepackage{graphicx}
\usepackage{epsfig}
\fi
\graphicspath{{./Figs/}{./}}

\usepackage{latexsym}
\usepackage{amstext}
\usepackage{amssymb}
\usepackage{bm}
\usepackage{wasysym}

\usepackage{hyperref}

\newcommand{\trc}{\mbox{trace}}

\newcommand{\mass}{\mathsf{m}}

\newcommand{\eexp}[1]{\mathrm{e}^{#1}}

\newcommand{\BraKet}[3]{ \left\langle #1 \middle| #2 \middle| #3 \right\rangle}

\newcommand{\be}[1]{\begin{eqnarray} \label{e#1}}
\newcommand{\beq}{\begin{eqnarray}}
\newcommand{\eeq}{\end{eqnarray}} 

\newcommand{\hide}[1]{}

\newcommand{\Eq}[1]  {{\textcolor{blue}{Eq.}}(\ref{#1})} 
\newcommand{\Fig}[1] {{\textcolor{blue}{Fig.}}\ref{#1}}

\newcommand{\hrefl}[1]{\href{#1}{[link]}}

\begin{document}


\title[Analysis of the AQUID]{Chaos and two-level dynamics of the Atomtronic Quantum Interference Device}

\author{Geva Arwas, Doron Cohen}

\address{
\mbox{Department of Physics, Ben-Gurion University, Beer-Sheva 84105, Israel} 
}

\begin{abstract}
We study the Atomtronics Quantum Interference Device employing a semiclassical perspective. 
We consider an $M$ site ring that is described by the Bose-Hubbard Hamiltonian.
Coherent Rabi oscillations in the flow of the current are feasible, 
with an enhanced frequency due to to chaos-assisted tunneling. 
We highlight the consequences of introducing a weak-link into the circuit. 
In the latter context we clarify the phase-space considerations that are involved in setting up an effective 
``systems plus bath" description in terms of Josephson-Caldeira-Leggett Hamiltonian.
\end{abstract}


\section{Introduction}

Atomtronics is a new quantum technology \cite{seaman2007atomtronics,pepino2009atomtronic,Amico-Boshier}, with potential for novel quantum computing implementations \cite{hallwood2006macroscopic,solenov2010metastable,Amico2014,Aghamalyan15}. Theory and experiments with Atomtronic superfluid circuits are in the focus of current research \cite{Amico_LG,NIST1,NIST2,Minguzzi}. A major objective is to realize a Quantum Interference Device (AQUID) that possibly includes one or two weak-links \cite{BoshirPRL2013}. This is analogous to a superconducting circuit, or to its low dimensional version (fluxon, Josephson vortex qubit) \cite{Hekking,Manucharyan}. However the design considerations of such device are still somewhat vague. 

We study an Atomtronic superfluid circuit that is described by the Bose-Hubbard Hamiltonian (BHH) \cite{Amico2014}. Namely, we consider $N$ bosons in an $M$~site rotating ring such that the model parameters are $(N,M,K,U,\Phi)$, where $K$ is the hopping frequency between the sites, $U$ is the on-site interaction, and the rotation is formally equivalent to having an Aharonov-Bohm flux~$\Phi$. If a weak-link is introduced, there is an additional parameter~$\alpha$ that characterizes the relative strength of the coupling.

For the purpose of qubit realization, the objective is to single out  
a two-level system (TLS) that is quasi-isolated from all the other microscopic degrees 
of freedom (DOFs). In the present context there are two flow-states that 
differ by their ``winding number" $m$, 
meaning that they are characterized by a different value 
of the persistent current ($I_m$).
The flow-states are required to be meta-stable, 
meaning that each of them will not decay in time. 
If they are quasi-degenerate,  
one would like to witness coherent Rabi oscillations.
During a Rabi-based protocol the system evolves into 
a superposition of macroscopically distinct 
flow-states \cite{leggett1980macroscopic}. 

The introduction of a weak-link allows control 
over the coupling $\Delta_s$ between the flow-states.
Without a weak-link this coupling  might be too small for operational purpose, 
meaning that the time period ($2\pi/\Delta_s$) 
of coherent Rabi oscillations might become too large
for practical implementations.  
The relative strength of the weak-link is characterized 
by a parameter $\alpha$. If ${\alpha<1}$ the weak link
destroys the meta-stability, which is effectively like having 
a disconnected ring. Hence one requires ${\alpha>1}$.   
The dependence of~$\Delta_s$ on~$N$ and on~$M$ in the case of an AQUID 
has been recently addressed in Ref.\cite{Aghamalyan15} following \cite{Minguzzi},
highlighting the subtle interplay of interactions and quantum fluctuations. 
The present work is in a sense complementary and provides 
a semi-classical perspective for the analysis of a few-site ring 
that is described by the BHH, with or without a weak-link.


Formally our BHH system has ${d=M{-}1}$ coupled DOFs:  
the dimer ($M=2$) is the so-called bosonic Josephson junction;  
while the trimer ($M=3$) is the minimal superfluid circuit.
Our main focus is on BHH circuits with ${M=3,4,5}$ sites, 
but we shall refer to ${M\ge 6}$ rings too.     
The following specific questions arise:
{\bf \ (A)}~In what range of the model parameters is it possible to have metastable flow-states?
{\bf \ (B)}~Can we treat two quasi-degenerate flow-states as a coherent two-level-system?
If yes, 
{\bf \ (C)}~how the frequency of the coherent Rabi oscillation is determined? 
And if a weak-link is introduces then, 
{\bf \ (D)}~can we derive the dynamics from an effective ``system plus bath" Hamiltonian.  
Question (A) has been partially addressed in our previous publications \cite{sfs,sfc}, 
and its physics is briefly summarized in \ref{a:sf}.  
In the present work we would like to further address questions (B-D).

Our main observations are: 
{\bf\ (1)}~In the absence of a weak-link, coherent Rabi oscillations 
are feasible, with frequency that is possibly determined by chaos-assistance tunneling, 
leading to weaker dependence on the number of particles. 
{\bf \ (2)}~In particular we demonstrate numerically Rabi oscillations 
between metastable flow-states in a non-rotating ($\Phi=0$) circuit that consists of $M=4$ sites. 
{\bf \ (3)}~We find what is the critical strength of a weak-link, 
below which superfluidity is diminished.  
{\bf \ (4)}~We illuminate how our considerations connect 
with the familiar ``system plus bath" framework of Caldeira and Leggett. 
{\bf \ (5)}~We show that with weak-link the threshold to chaos 
is pushed up in energy, which is a necessary condition for the validity  
of the single Josephson-junction description. 
{\bf \ (6)}~We point out that the requirement for observing 
coherent Rabi oscillation in large $M$ rings might be in clash 
with the quantum Mott transition.

{\em The outline is as follows:} 
In Section~2 we introduce the model and the methods; 
In Section~3 we discuss the coherent dynamics in the absence of a weak-link.
In Sections~4 and~5 we analyze how a weak-link affects 
a ring with {\em few} or {\em many} sites respectively. 
We care to make a bridge between the semiclassical and the ``system plus bath" perspectives. 
Finally we summarize the overall picture in Section~6.

\section{Model and Methods}

We consider $N$ Bosons in an $M$~site rotating ring such that the model parameters are $(N,M,K,U,\Phi)$, 
where $K$ is the hopping frequency between the sites, $U$ is the on-site interaction,
and the rotation is formally equivalent to having an Aharonov-Bohm flux~$\Phi$. 
If a weak-link is introduced, there is an additional 
parameter $\alpha \sim K^{\prime}/K$ that characterizes the relative strength 
of the coupling. Accordingly the ring is described by the Bose-Hubbard Hamiltonian (BHH):
\be{1}
\mathcal{H}_{\text{BHH}} \ \ = \ \ \sum_{j=1}^{M} \left[
\frac{U}{2} \bm{n}_{j}(\bm{n}_{j}-1) 
- \frac{K_j}{2} \left(\eexp{i(\Phi/M)} \bm{a}_{j{+}1}^{\dag} \bm{a}_{j} + \text{h.c.} \right)
\right]~.
\eeq
where $\bm{a}_j \, (\bm{a}^{\dagger}_j)$ are bosonic annihilation (creation) operators on the $j$th site
and $n_{j}= \bm{a}_{j}^{\dag}\bm{a}_{j}$ is the corresponding number operator.
Periodic boundaries are imposed, meaning that $\bm{a}_{M} \equiv \bm{a}_0$.
The parameter $U$ takes into account the finite scattering length for the atomic 
two-body collisions on the same site. The  hopping parameters are constant $K_j=K$ 
except in the weak-link where it is $K^{\prime}$.   
The ring is pierced by an artificial (dimensionless) magnetic flux $\Phi$,
which can be experimentally induced for neutral atoms as a Coriolis flux by rotating the lattice
at constant velocity~\cite{fetter,wright}, 
or as a synthetic gauge flux by imparting a geometric phase
directly to the atoms via suitably designed laser fields~\cite{berry,synth1,dalibard}.
The presence of the flux $\Phi$ in \Eq{e1} has been taken into account through 
the Peierls substitution: $K_j \rightarrow e^{-i(\Phi/M)} K_j$.

In the quantum analysis, we diagonalize \Eq{e1}, and display the spectrum as in \Fig{fg_bloch1}a. 
For each eigenstate $E_{\alpha}$ we calculate the fragmentation measure $\mathcal{M}$ as defined in \ref{a:fm}, 
while the average current is obtained using the following formula:
\beq
I_{\alpha} \ \ = \ \ \BraKet{E_{\alpha}}{-\frac{\partial \mathcal{H}}{\partial \Phi}}{E_{\alpha}}
\eeq
In a classical context the average is taken over time for a very long trajectory.

\subsection{Semiclassical perspective}

For the purpose of semiclassical analysis it is convenient 
to write the BHH using action-angle variables: $\bm{a} \mapsto \sqrt{\bm{n}} e^{i\bm{\varphi}}$. 
Accordingly the Hamiltonian describes an $M$ degrees of freedom (DOFs) system, namely,  
\be{2}
H \ \ = \ \ 
\sum_{j=1}^{M} \left[\frac{U}{2} \bm{n}_{j}^2 
-K_j \sqrt{\bm{n}_{j{+}1} \bm{n}_{j}} \, \cos\left(\bm{\varphi}_{j{+}1}{-}\bm{\varphi}_{j} - \frac{\Phi}{M} \right)
\right]   
\eeq
Since the total number of particles $N=\sum \bm{n}_j$ is a constant of the motion of the system, 
the Hamiltonian above describes ${d=M{-}1}$ coupled pendula: 
the dimer ($M=2$) is the so-called bosonic Josephson junction;  
while the trimer ($M=3$) is the minimal superfluid circuit.
Our main focus is on BHH circuits with ${M=3,4,5}$ sites, 
but we shall refer to ${M \ge 6}$ rings too.  
The interaction is characterized by the dimensionless parameter
\beq
u \ \ = \ \ \frac{NU}{K} 
\eeq
The classical dynamics is governed by
\be{100}
\dot{z} = \mathbb{J} \, \bm{\partial} H, 
\ \ \ \ \ \ \ \ \ \ \ \ 
\mathbb{J} \equiv  
\left( 
\scriptscriptstyle{
\begin{array}{cccc}
0 & \mathbb{I} \\
-\mathbb{I} & 0 \\
\end{array}
}
\right) 
\eeq
where $z \equiv (\bm{\varphi}_1,\cdots,\bm{\varphi}_M,\bm{n}_1,\cdots,\bm{n}_M)$ 
are the canonical coordinates.
The notation $\partial_{\nu}$ stands for derivative with respect to $z_{\nu}$, 
and $\mathbb{J}$ is the symplectic matrix. 
It is important to emphasize that upon re-scaling the only dimensionless 
parameters that affect the classical trajectories are $(u,\Phi)$ and $K^{\prime}/K$.  
The effective Planck constant is $\hbar=1/N$. The latter parameter, 
does not appear in the ``classical" equations of motion \Eq{e100}, 
but only in the full quantum treatment of \Eq{e1}.

\subsection{System plus bath perspective}

The conventional approach for analyzing a SQUID/AQUID is based on a ``system plus bath" perspective. 
This perspective becomes meaningful once a weak-link is introduced, 
which is like having a ``slow DOF".   
In order to motivate the conventional phenomenology one can 
regard the BHH \Eq{e2} as describing masses that are connected
by nonlinear springs. If one spring is very ``weak",
then at low energies the equal-partition theorem justifies 
an harmonic approximation for the small vibrations of the 
other springs. Accordingly we can regard the system has having 
one non-linear DOF ("pendulum") coupled to phonons ("harmonic bath").
The canonical coordinates that describe the weak-link 
are the phase difference ${\bm{\varphi} = (\bm{\varphi}_M - \bm{\varphi}_1)}$, 
and its conjugate ${\bm{n}=(\bm{n}_M - \bm{n}_1)/2}$. 
Hence we obtain the the Josephson Circuit Hamiltonian (JCH) 
\be{5}
\mathcal{H}_{\text{JCH}} \ \ = \ \ E_C \ \bm{n}^2 
+ \frac{1}{2}E_L \bm{\varphi}^2
- E_J \ \cos(\bm{\varphi}-\Phi)  
+ \mathcal{H}_{\text{bath}}
\eeq
with ${E_C=U}$, and $E_L=[(N/M)/(M-1)]K$, and $E_J=(N/M)K^{\prime}$. 
The bath Hamiltonian has the standard Caldeira-Leggett form
\be{3}
\mathcal{H}_{\text{bath}} \ \ = \ \  
\sum_{m} 
\left( \frac{1}{2\mass_m}	\tilde{\bm{n}}_{m}^2  
+ \frac{1}{2}\mass_m\omega_m^2  
\left( \tilde{\bm{\varphi}}_m - \frac{c_m}{\mass_m\omega_m^2} \bm{\varphi}  \right)^2
\right)
\eeq
For small $M$ the ``bath" is merely a set of several oscillators, 
and possibly can be neglected, because the $\omega_m$ are typically large
compared with the natural frequency of the junction.  
For large $M$ one can characterize the bath oscillators 
by an Ohmic spectral function
\beq
J(\omega) \ \ \equiv \ \      
\frac{\pi}{2} \sum_m \frac{c_m^2}{\mass_m \omega_m} \delta(\omega-\omega_m) 
\ \ = \ \ \eta \omega \ (\omega<\omega_c), 
\eeq
The detailed derivation and the explicit expressions 
for the bath parameters in terms of the BHH parameters 
are presented in \ref{a:jch}, 
and will be further discussed in a later section. 
We note that in \cite{Hekking,Amico2014} the finite-temperature partition-function 
of the BHH ring has been introduced, and the reduced ``system plus bath" action has been deduced. 
From the reduced action one could figure-out what is the effective JCH.
In the present  approach to the {\em same} system, we do not assume finite temperature, 
but merely re-arrange the Hamiltonian in a way that allows a ``system plus bath"   
description. This is a valid procedure even if the ring is prepared (say) 
in a micro-canonical state with some arbitrary energy~$E$. 
One may say that in our treatment $E/M$ plays the role of the temperature.

Within the framework of the JCH treatment, the possibility of having
metastable flow-states is controlled by the parameter
\be{9}
\alpha \ \ \equiv \ \ \frac{E_J}{E_L} \ \ = \ \ (M-1) \ \frac{K^{\prime}}{K}
\eeq
For ${\Phi=\pi}$ the condition for having at least two local minima in the potential floor of the JCH, 
is ${\alpha>\alpha_c}$, where ${\alpha_c=1}$.
Disregarding small quantum fluctuations, the two local minima can support 
a quasi-degenerate pair of flow-states.
If the bath is ignored, then from the WKB approximation 
it follows that the tunnel splitting is given by 
some variation of the following expression \cite{Hekking}  
%
\be{90}
\Delta_s \ \ \approx \ \ \sqrt{E_C E_J} \ 
\exp \left[ - C \sqrt{\frac{E_J}{E_C}} \right] 
%
%
\eeq  
where $C$ is a numerical prefactor. We would like to emphasize that there are several 
variations of this formula, depending on the relative size of $(E_C,E_L,E_J)$, 
but they are all based on the assumption that \Eq{e5} is a valid description.

\subsection{Two-level system perspective}

The objective is obviously to realize a two-level system (TLS) 
that is quasi-isolated from all the other microscopic DOFs  
\cite{Rey07,Nunnenkamp08,Hallwood06,Hallwood09,Nunnenkamp11,Nunnenkamp10,Aghamalyan15}. 
In the present context there are two flow-states 
that differ by their ``winding number" $m$, 
meaning that they are characterized by a different value 
of the persistent current ($I_m$). 
We label these states as $\circlearrowright$ and $\circlearrowleft$, 
and write the TLS hamiltonian as 
\be{10}
\mathcal{H}_{\text{TLS}} \ \ = \ \ 
\left(\matrix{E_{\circlearrowright} & \Delta_s/2 \cr \Delta_s/2 &  E_{\circlearrowleft}}\right)
\eeq
We refer to $\Delta_s$ as the splitting: if we draw the eigen-energies  
versus $\Phi$ we get an avoided crossing. 
The flow-states are required to be meta-stable, 
meaning that each of them will not decay in time. 
If they are quasi-degenerate,  
one would like to witness coherent Rabi oscillations.
The quasi-degeneracy is controlled by~$\Phi$, 
and happens for $\Phi=0$ (say $m=\pm1$) 
or for $\Phi=\pi$ (say $m=0,1$).          
During the Rabi oscillation the system evolves into 
a superposition of these macroscopically distinct flow-states. 
Such superposition is commonly termed ``cat state".  
  
The conventional procedure to engineer a TLS is as follows:
{\bf (i)}~To introduce a ring with a weak-link; 
{\bf (ii)}~To ensure that the weak-link DOF is  
only weakly-coupled to all the other ring DOFs;
{\bf (iii)}~To analyze the operation of the device using 
the ``system plus bath" paradigm of Caldeira and Leggett.
The introduction of a weak-link allows the reduction 
of the many-body BHH \Eq{e1} into the simpler JCH \Eq{e5}.
The JCH consists of a single pendulum-like DOF 
that interacts weakly with harmonic-oscillators ("phonons").
The relative strength of the weak-link is characterized 
by a parameter $\alpha$ of \Eq{e9}. If ${\alpha<\alpha_c}$ the weak-link
destroys the meta-stability (it is effectively like having 
a disconnected ring), hence we require ${\alpha>\alpha_c}$.   
Furthermore, the introduction of a weak-link allows control 
over the coupling $\Delta_s$ between the flow-states. 
Without a weak-link this coupling  might be too small for operational purpose, 
meaning that the time period ($2\pi/\Delta_s$) 
of coherent Rabi oscillations might become too large
for practical implementations.  

If the bath is taken into account then there are two effects. 
One is ``dressing" of the bare parameters, and the other is ``noise". 
It is well known from the work of Caldeira and Leggett
that coherent Rabi oscillations can be observed provided ${\eta<\eta_c}$, 
where $\eta_c$ is of order unity (${\eta_c=\pi}$ for the spin-boson model).
We shall come back to this issue when we discuss the large $M$ limit.

\section{Coherent dynamics in the absence of a weak-link}

\begin{figure}
\begin{center}

\includegraphics[width=0.45\hsize]{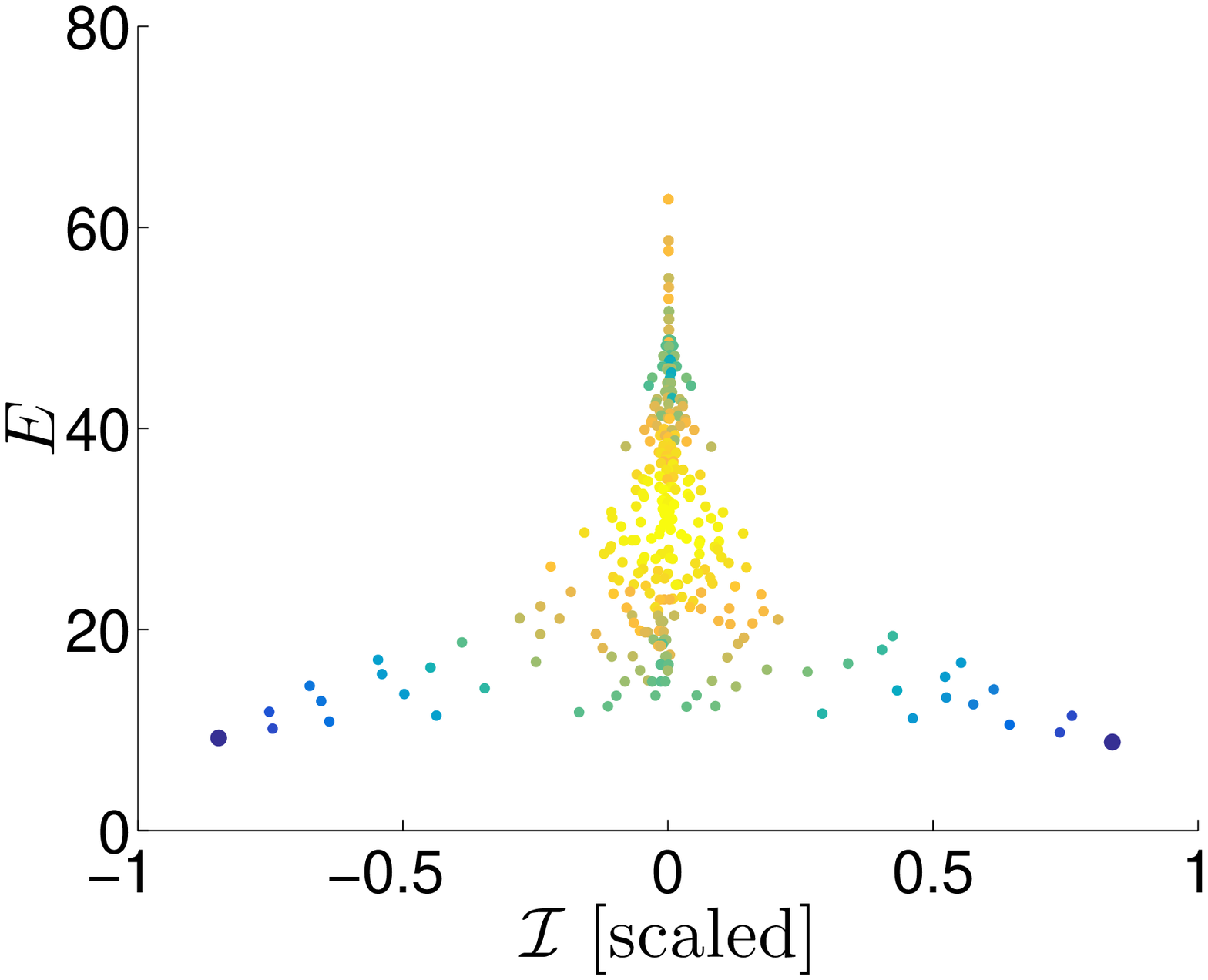} 
\includegraphics[width=0.45\hsize]{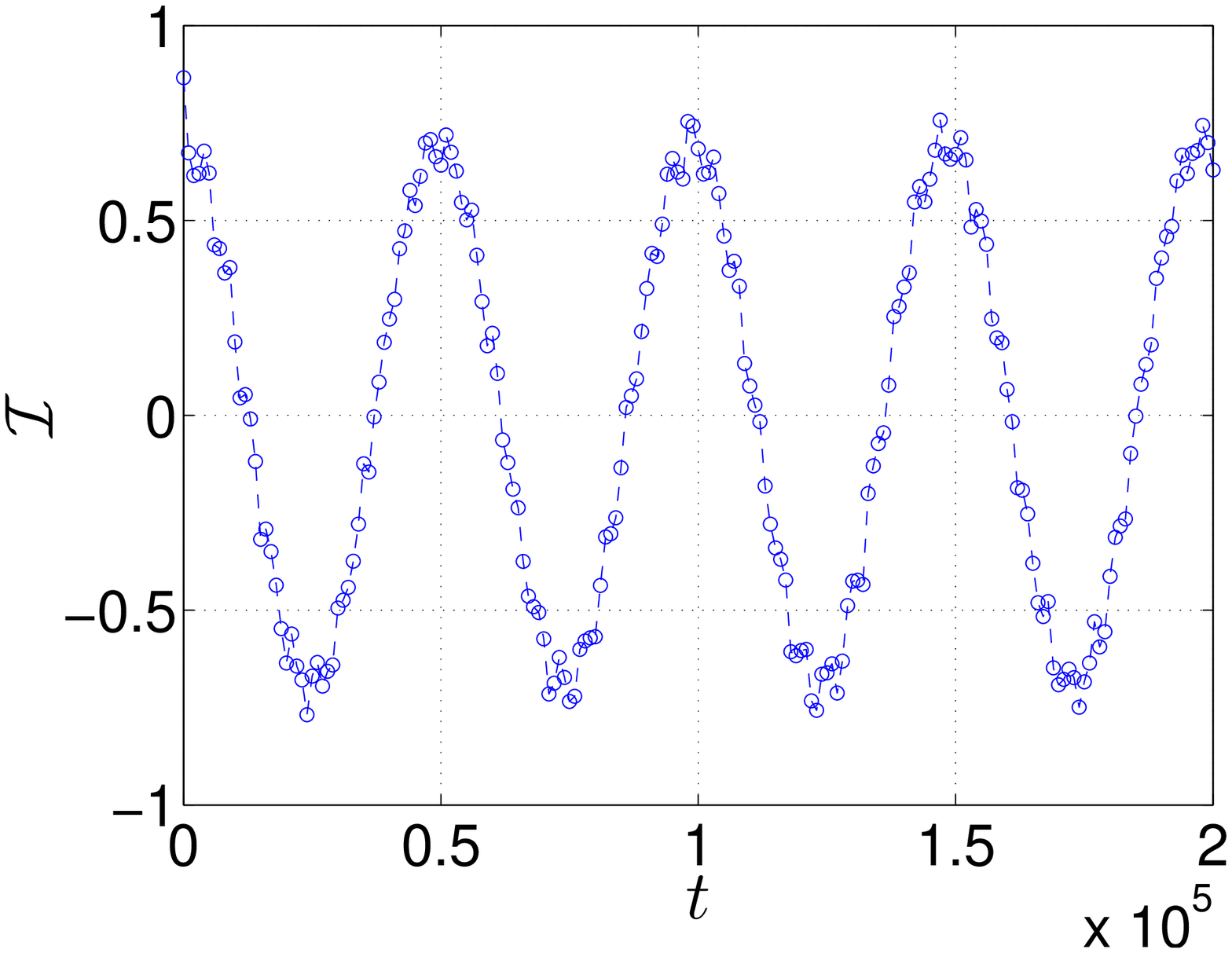} 

\caption{ \label{fg_bloch1} 
Spectrum of $M{=}3$ ring with $N{=}24$ bosons, interaction $u{=}5$, and $\Phi{\sim}\pi$ rotation (left panel,~a);   
accompanied with simulation of Rabi oscillations for $\Phi=\pi$ rotation (right panel,~b). 
The units of time (here and in the subsequent figures) are fixed by the hopping frequency~$K=1$.
In (a) each point represents an eigenstate, positioned according to its energy $E_{\alpha}$ (vertical axis) 
and its current $I_{\alpha}$ (horizontal axis). The current is in units of $NK/M$. 
The color encodes the fragmentation of each eigenstate (blue $\mathcal{M} \sim1$ to red $\mathcal{M}\sim M$).
The quasi-degenerate flow-states at the bottom of the energy landscape 
are energetically-stable (``Landau stability") and are separated by a forbidden-region.
The tunnel-coupling allows coherent Rabi oscillations with extremely slow frequency~$\Delta_s$. 
If we did not slightly perturbed~$\Phi$, the diagonalization would give 
zero current cat-states (symmetric and anti-symmetric superposition of the pertinent flow-states).  
In (b) the initial state is an $m=1$ coherent state, and the system has exactly $\Phi{=}\pi$ rotation.
This initial state has large overlap with the pair of quasi-degenerate cat eigenstates.
Consequently we observe Rabi oscillations of the current with frequency $\Delta_s$
that is determined by the tunnel coupling.   
} 
 
\ \\

\includegraphics[width=0.45\hsize]{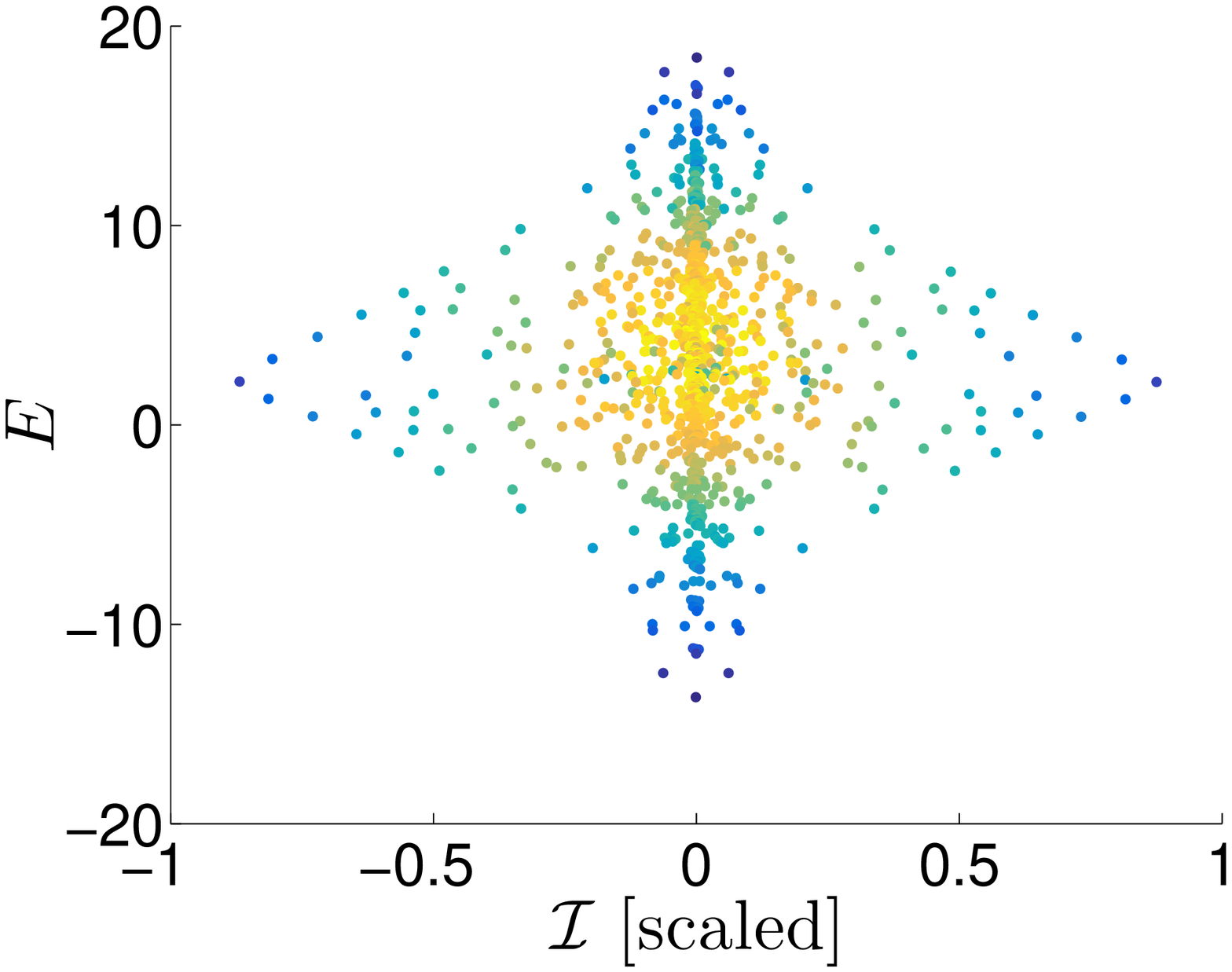} 
\includegraphics[width=0.45\hsize]{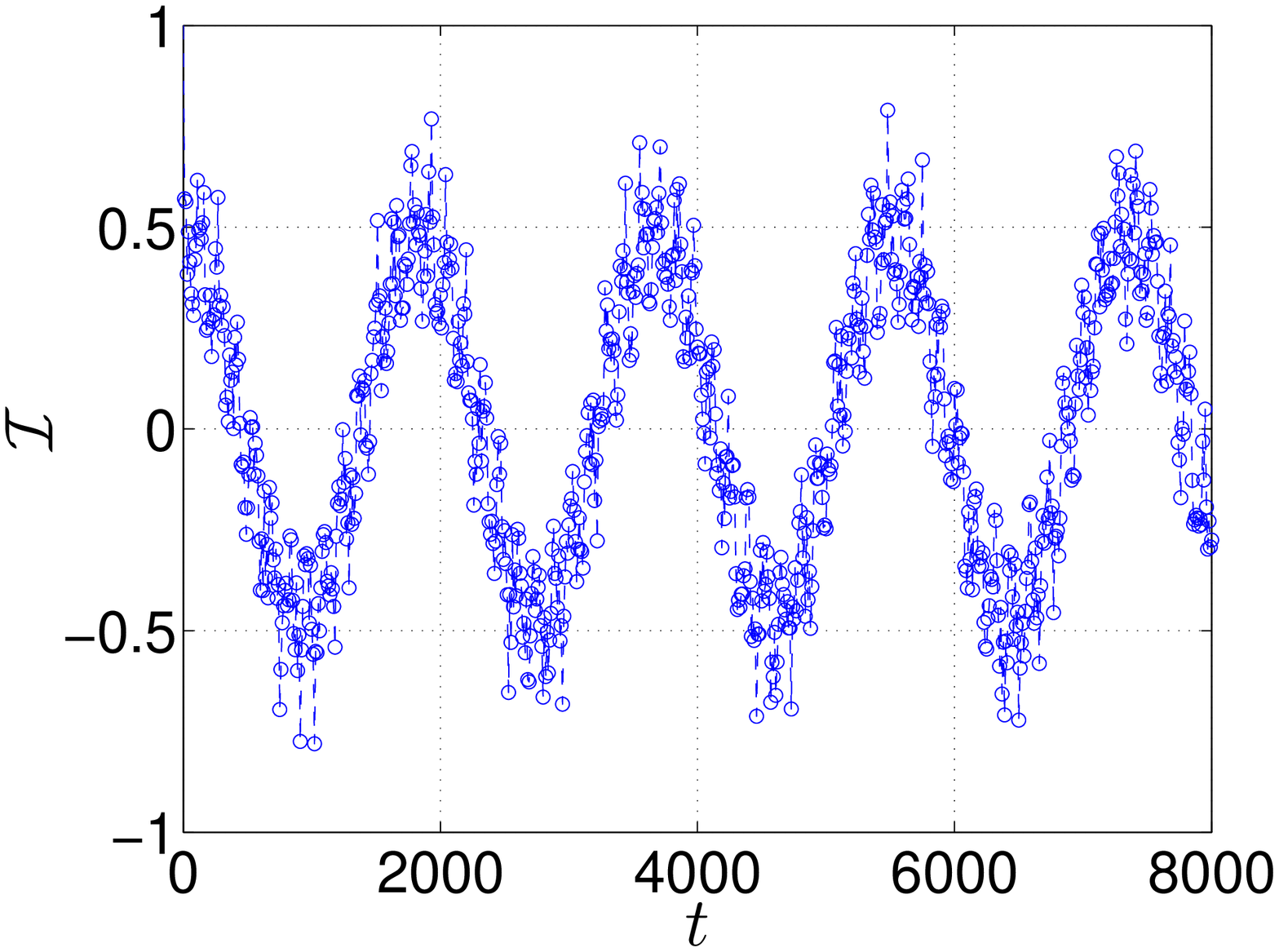} 

\caption{ \label{fg_bloch2} 
Spectrum of $M{=}4$ ring with $N{=}16$ bosons, interaction $u{=}1$ and $\Phi \sim 0$ rotation (left panel,~a); 
accompanied with simulation of Rabi oscillations (right panel~,b).  
Here the quantum meta-stability of the flow-states $m=\pm1$ 
is related to quantum localization on an Arnold web.  
The coupling is mediated by a chaotic sea. 
Consequently we observe chaos-assisted coherent Rabi oscillations 
with relatively short period, 
which is important for practical qubit implementation. 
} 
 
\end{center}
\end{figure}

The stationary orbitals of a {\em single} particle in a clean ring 
are the momentum states with wavenumber $k=(2\pi/M)m$, 
where $m$ is an integer modulo~$M$.
Coherent flow-states have $N$ particles {\em condensed}  
into the same momentum orbital: 
\beq 
| m \rangle \ \ \equiv \ \ \left( \tilde{\bm{a}}_m^{\dagger}  \right)^N | 0 \rangle 
\eeq
Implying a macroscipically large current
\beq\label{eq:current}
\mathcal{I}_m \ \ = \ \ N \times \left( \frac{K}{M}\right) \sin  \left( \frac{1}{M}(2\pi m-\Phi) \right)
\eeq
In the absence of interaction ($U=0$) these coherent flow-states
are the eigenstates of the BHH. For $\Phi=\pi$ the $m=0$ and $m=1$ 
flow-states are degenerate in energy. 
If we add not-too-strong interaction they become coupled 
and may form a doublet whose dynamics is generated by the TLS hamiltonian \Eq{e10}.
The energy-difference ${\delta E \equiv E_{\circlearrowright}-E_{\circlearrowleft}}$  
is determined by the deviation $\delta\Phi \equiv (\Phi - \pi)$, 
and the coupling $\Delta_s$ is determined
by the strength of the interaction. 
An example for such doublet if provided in \Fig{fg_bloch1}.

Assuming that we have a TLS doublet of flow-states with energy splitting $\Delta_s$, 
one would expect to witness pure Rabi oscillations. 
If the system has been prepared (say) in a flow-state with clockwise current, 
the subsequent evolution would be  
\beq
| \Psi(t) \rangle \ \ = \ \ \cos\left(\frac{\Delta_s t}{2}\right) | \circlearrowright \rangle \ - i \sin\left(\frac{\Delta_s t}{2}\right) |  \circlearrowleft \rangle 
\eeq
implying alternating current with frequency $\Delta_s$, namely, 
\beq
\left\langle \mathcal{I}(t) \right\rangle \ \ = \ \  
\cos^2\left(\frac{\Delta_s t}{2}\right) \mathcal{I}_{\circlearrowright} \ + \sin^2\left(\frac{\Delta_s t}{2}\right) \mathcal{I}_{\circlearrowleft} 
\eeq

If we add weak-link or weak-disorder, 
the flow-states remain stable, provided the perturbation 
is not strong compared with the interaction.    
This is the essence of superfluidity.  
The stability is due to the non-zero interaction~$U$.
The interaction stabilizes the flow-states: 
instead of being located on a flat potential floor, 
the flow-states are located in local minima of the potential floor.
Local minima are structurally-stable with respect 
to the added disorder, i.e. the local minima 
do not diminished by a weak perturbation. 
The common conception is that the the two minima
are separated by a ``forbidden region". 
This is the same reasoning  that leads to \Eq{e90}, 
but here we refer to the multi-dimensional phase-space of the BHH \Eq{e2} 
and not to the reduced single DOF description of \Eq{e5}.  
Nevertheless, both perspectives connect smoothly.
Namely, a rough way to write \Eq{e90}, that illuminates the semiclassics is  
\beq
\Delta_s \ \ \sim \ \  
\Delta_0 \ \exp\left[ - C_M \ N \ \sqrt{\frac{\alpha}{u}} \right]  
\eeq  
where the prefactor $C_M$ has some dependence on~$M$.
This version highlights the distinction between 
the ``classical" parameters $(u,\alpha)$ and the ``quantum" parameter $\hbar=1/N$. 
In the absence of a weak-link one formally makes 
the sunstitution $\alpha \mapsto (M{-}1)$ as implied by \Eq{e9}.   
The energy scale $\Delta_0 \equiv (E_LE_C)^{1/2}$ 
is like the ``attempt frequency" of the Gamow-formula.
In a later section we identify $\Delta_0$  
as the frequency spacing between the phononic modes. 
   
In the JCH based picture, the splitting $\Delta_s$ is exponentially small in $N$   
due to the existence of a classically ``forbidden region" 
between the two local minima,  which necessitates {\em tunneling}.   
This very small $\Delta_s$ creates difficulties in witnessing coherent two-level dynamics in such configuration.
In order to have a bigger $\Delta_s$ a smaller $\alpha$ is required. 
But is should not be smaller than $\alpha_c=1$ else the meta-stability is diminished.
Note also that there is a trade-off between the weakness of the link 
and the quality of the superposition state \cite{Nunnenkamp11}.

The question arises whether one can manage without introducing a weak-link. 
In fact there is a loophole. In order to realize this loophole, 
one should be aware, following \cite{sfc},  
that there are novel flow-states that are not supported by 
local minima of the potential, but by a {\em ``stability island"}
or by a {\em ``chaotic pond"}, or by an {\em ``Arnold web"} region. 
We summarize all these possibilities in \ref{a:sf} - the exact details are not important. 
The important point is that the phase-space locations,  
where the flow-states reside, are not separated  by a {\em ``forbidden region"}.
Instead they are separated  by a {\em ``chaotic-sea"}.   
A visualization of this possibility is provided by the quantum spectrum in \Fig{fg_bloch2}, 
which should be contrasted with that of \Fig{fg_bloch1}. 
The way we plot the quantum spectrum (following \cite{sfc}) 
is in one-to-one correspondence with a section of the classical phase-space:  
In \Fig{fg_bloch1} the two flow-states at the bottom are 
separated by a ``forbidden region" where no states can reside; 
In contrast to that, in \Fig{fg_bloch2}, between the two metastable states 
there are many other states with roughly the same energy that reside in the ``chaotic-sea".  
 
If the coupling between the quasi-degenerate 
eigenstates is mediated by a chaotic sea, 
then $\Delta_s$ is much larger.
This is known as chaos-assisted tunneling \cite{CAtunneling1,CAtunneling2,CAtunneling3,CAtunnelingMoiseyev}.
Possibly the term tunneling is not the best description 
for the mathematics that is involved. The rough idea is that 
the quantum-coupling between the two metastable states   
is mediated by some intermediate state in the chaotic sea.
The coupling is roughly estimated using second-order 
perturbation theory as $\Delta_s \sim U^2/\Delta$, 
where $\Delta$ is the detuning from exact resonance. 
This expression does not contain a WKB suppression exponent, 
so it is not small, but nevertheless it is very sensitive 
to the model parameters, as in the theory of universal conductance fluctuations.
   
\begin{figure}
\begin{center}

\includegraphics[width=0.45\hsize]{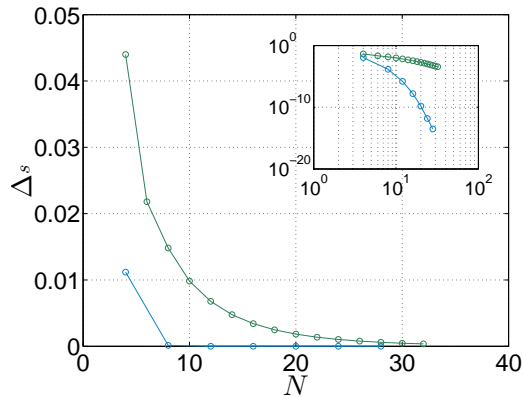} 

\caption{ \label{fg_bloch3} 
The frequency of the Rabi oscillations $\Delta_s$ is plotted     
as a function of the number of particles~$N$, for an $M=4$ site ring. 
The ``classical" parameter ${u=1}$ is kept constant.    
The lower curve is the $\Delta_s$ for oscillations between $m=0$ and $m=1$ at $\Phi=\pi$. 
The upper curve is the $\Delta_s$ for oscillations between $m=+1$ and $m=-1$ for $\Phi=0$.
The large $\Delta_s$ in the latter case is due to chaos-assisted tunneling.  
} 

\end{center}
\end{figure}

In \Fig{fg_bloch2} we provide a numerical demonstration of chaos-assisted Rabi oscillations.
In this example the device is non-rotating ($\Phi{=}0$), and the Rabi oscillations 
are between the metastable $m=\pm1$ flow-states. 
The dependence of $\Delta_s$ on the number of particles 
for ``chaos assisted tunneling" is contrasted with ``under the barrier tunneling" in \Fig{fg_bloch3}.   
 
Summarizing this section, we observe that the coupling between metastable flow-states can 
be via chaos-assisted tunneling, implying a relatively large $\Delta_s$ 
when compared with the conventional expectation. 
A weak-link in a few-site ring is not essential for getting large~$\Delta_s$. 
In fact its introduction is likely to be harmful for the device operation (see next section).

\section{weak-link in a few site ring}

\begin{figure}
\begin{center}

\includegraphics[width=0.45\hsize]{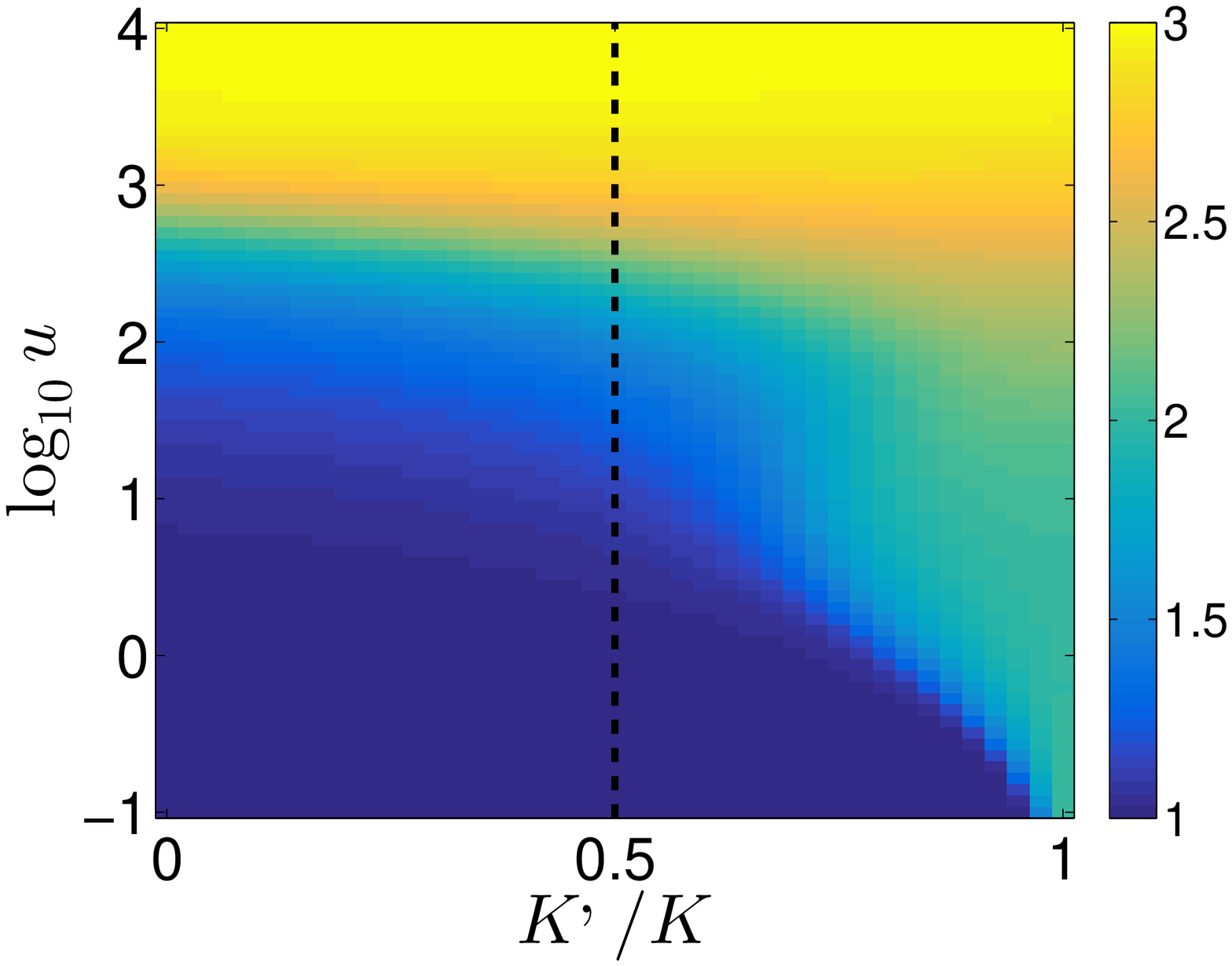}
\includegraphics[width=0.45\hsize]{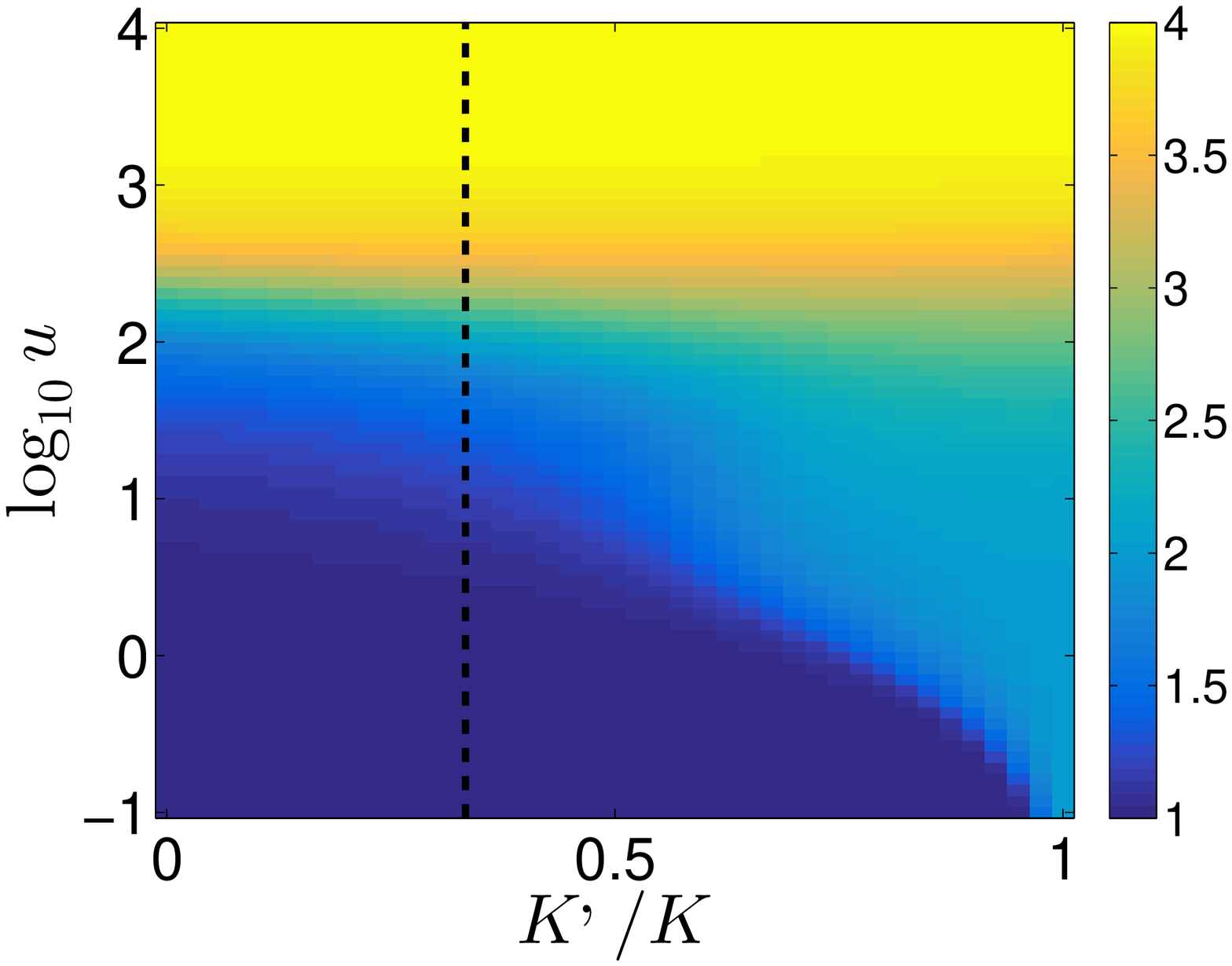} 

\caption{ \label{fig1}  
The fragmentation ($\mathcal{M}$) of the ground state is imaged as a function 
of $u$ and $K'/K$ for $M{=}3$ ring with $N{=}30$ particles (left) 
and for $M{=}4$ ring with $N{=}20$ particles (right). 
The value $\mathcal{M}=1$ indicates a coherent state (all particles are condensed 
in a single orbital). The value of ${\mathcal{M} \sim 2}$ indicates quasi degeneracy 
of the ground state (a doublet of flow-states). The value ${\mathcal{M} \sim M}$ 
indicates a fragmented state: here it is due to the quantum Mott transition.  
The vertical dashed line corresponds to the ${\alpha_c=1}$ border, 
which in the absence of a Mott transition would become valid for large~$u$.  
} 

\ \\

\includegraphics[width=0.24\hsize]{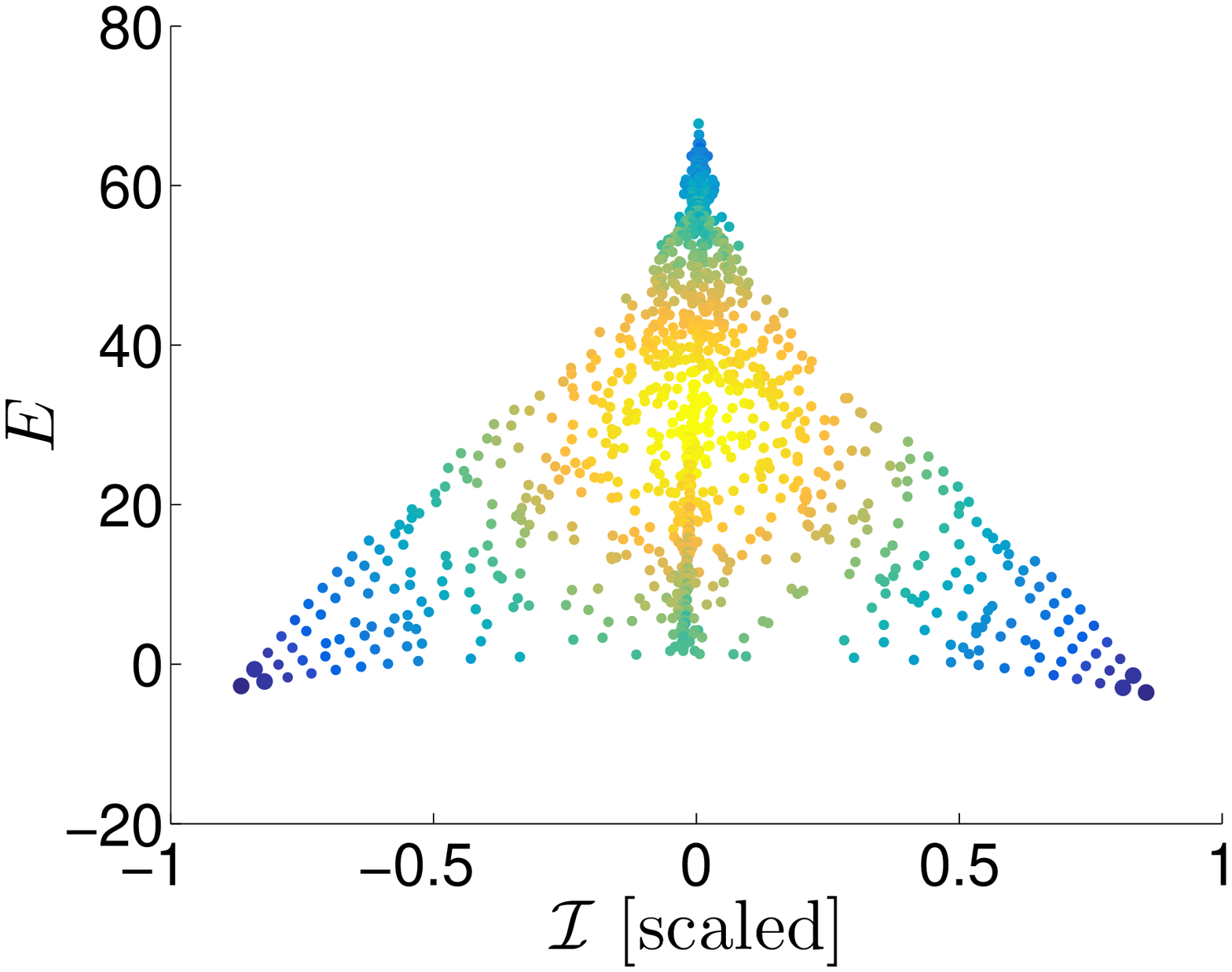} 
\includegraphics[width=0.24\hsize]{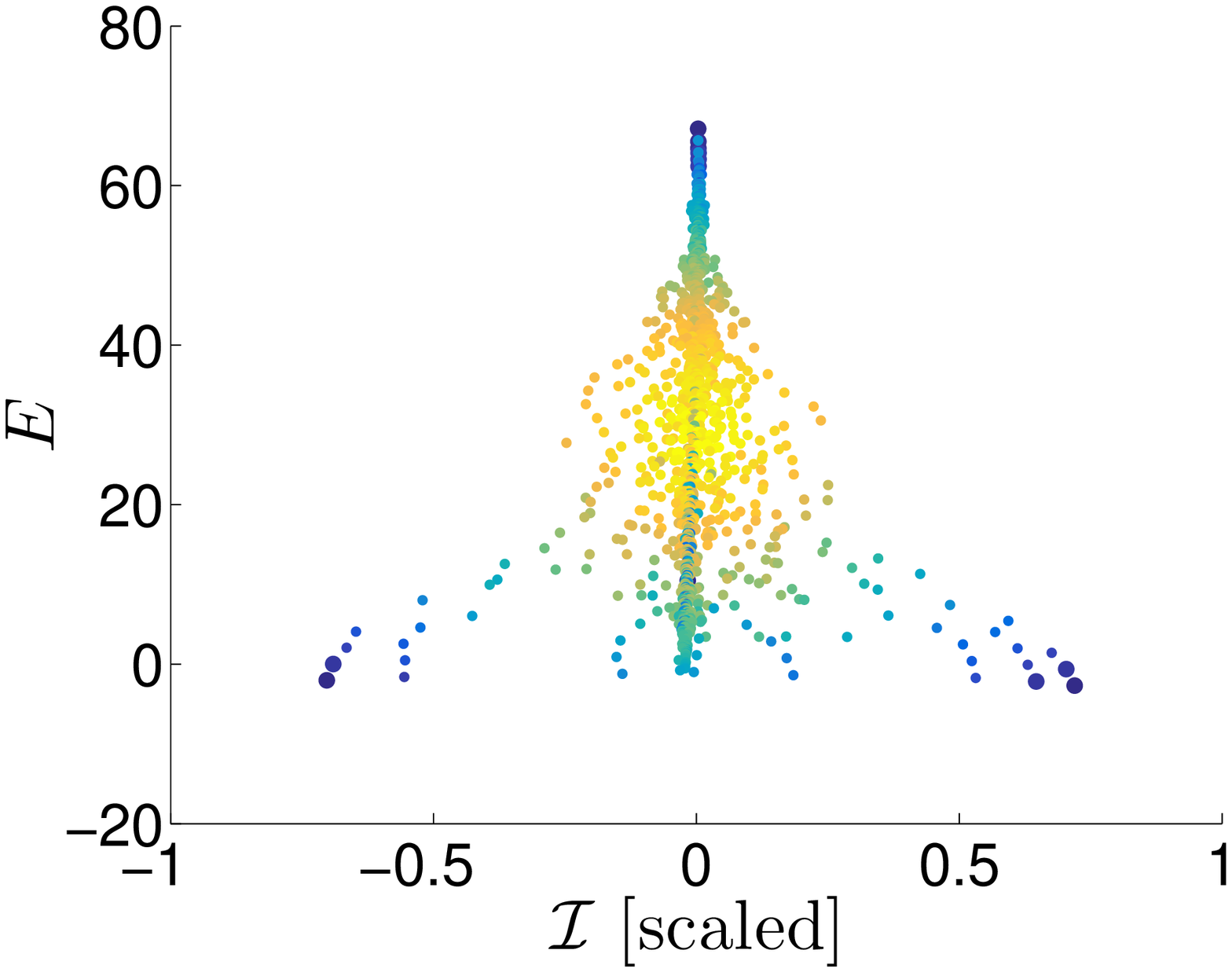} 
\includegraphics[width=0.24\hsize]{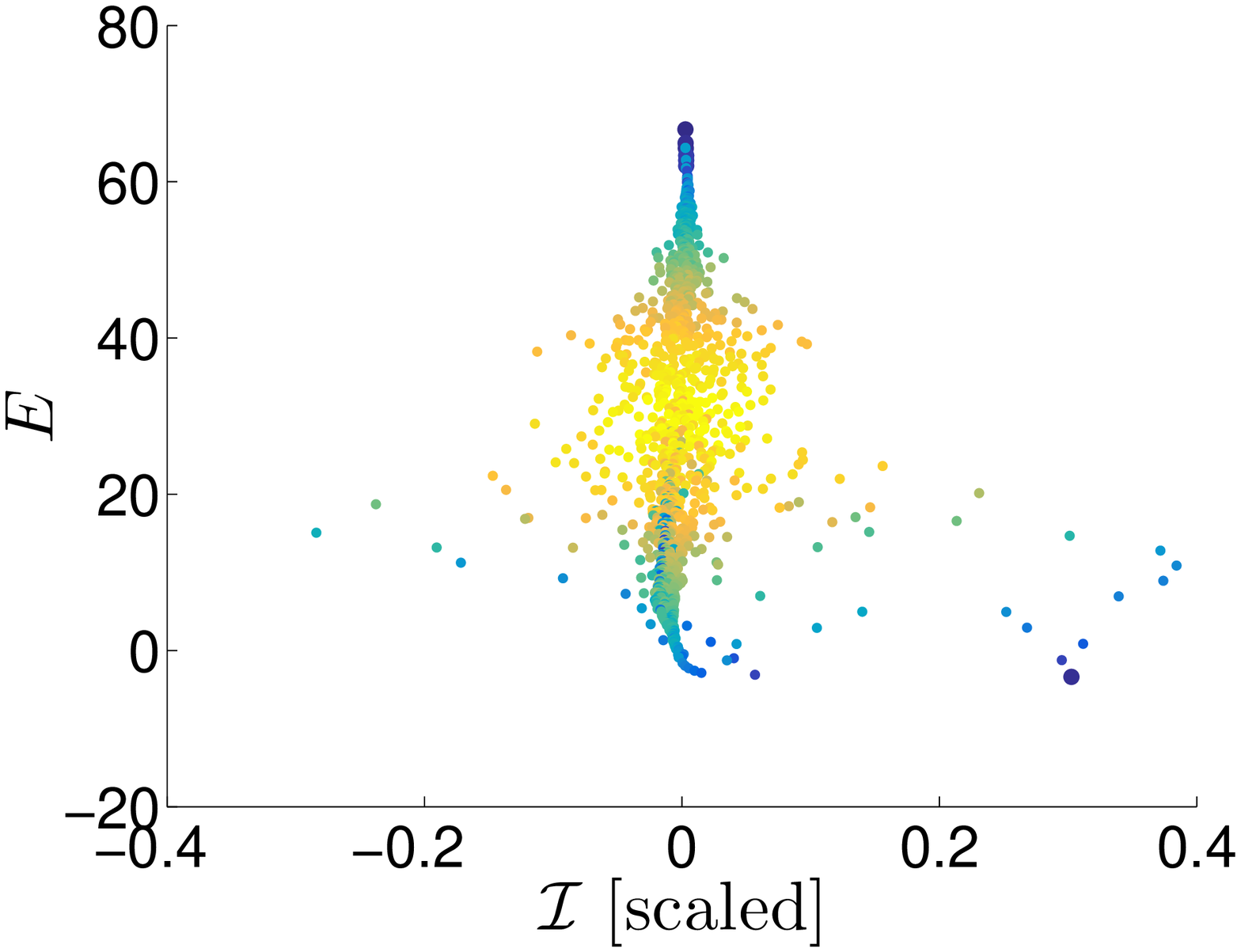} 
\includegraphics[width=0.24\hsize]{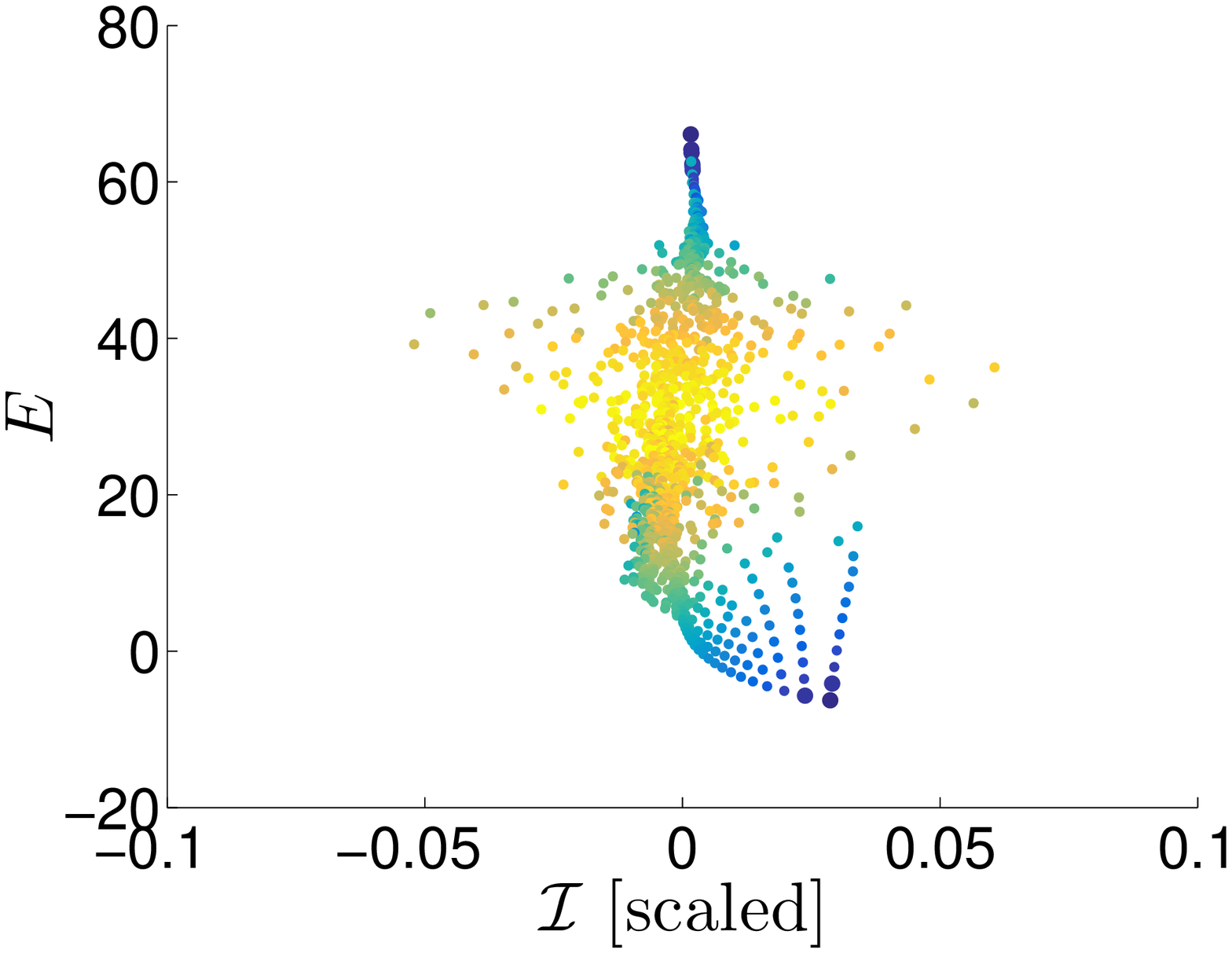}  
\\ \ \\ 
\includegraphics[width=0.24\hsize]{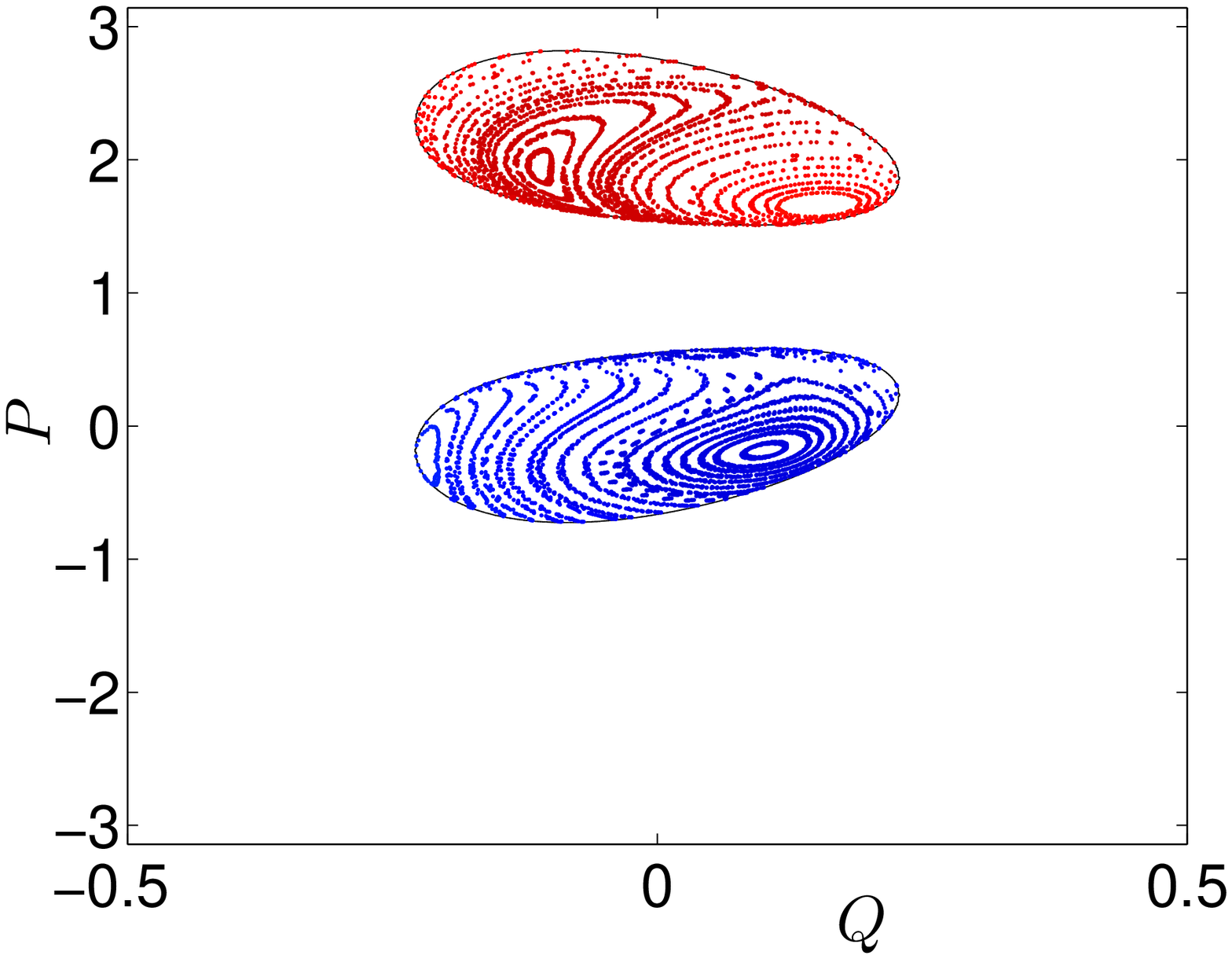} 
\includegraphics[width=0.24\hsize]{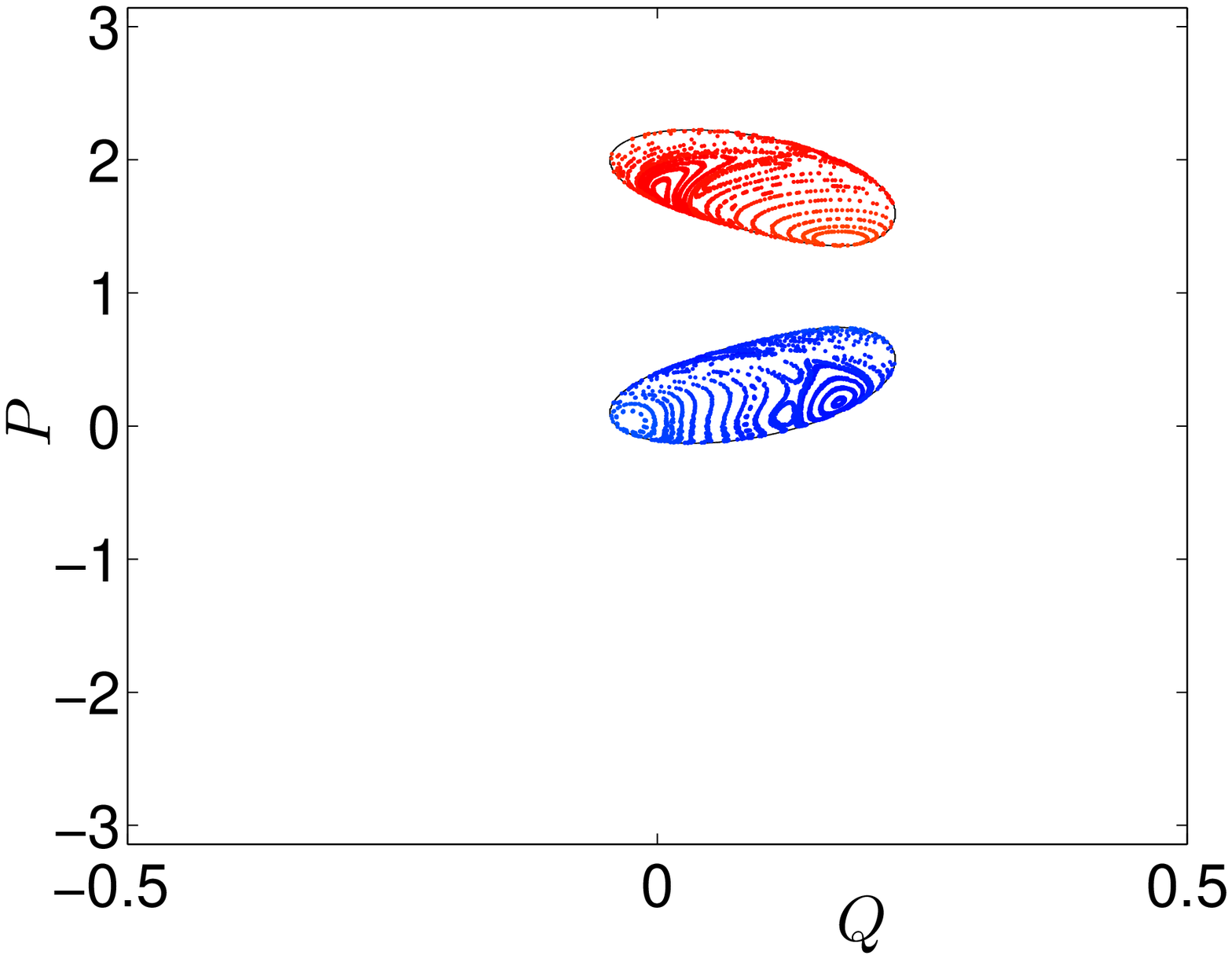} 
\includegraphics[width=0.24\hsize]{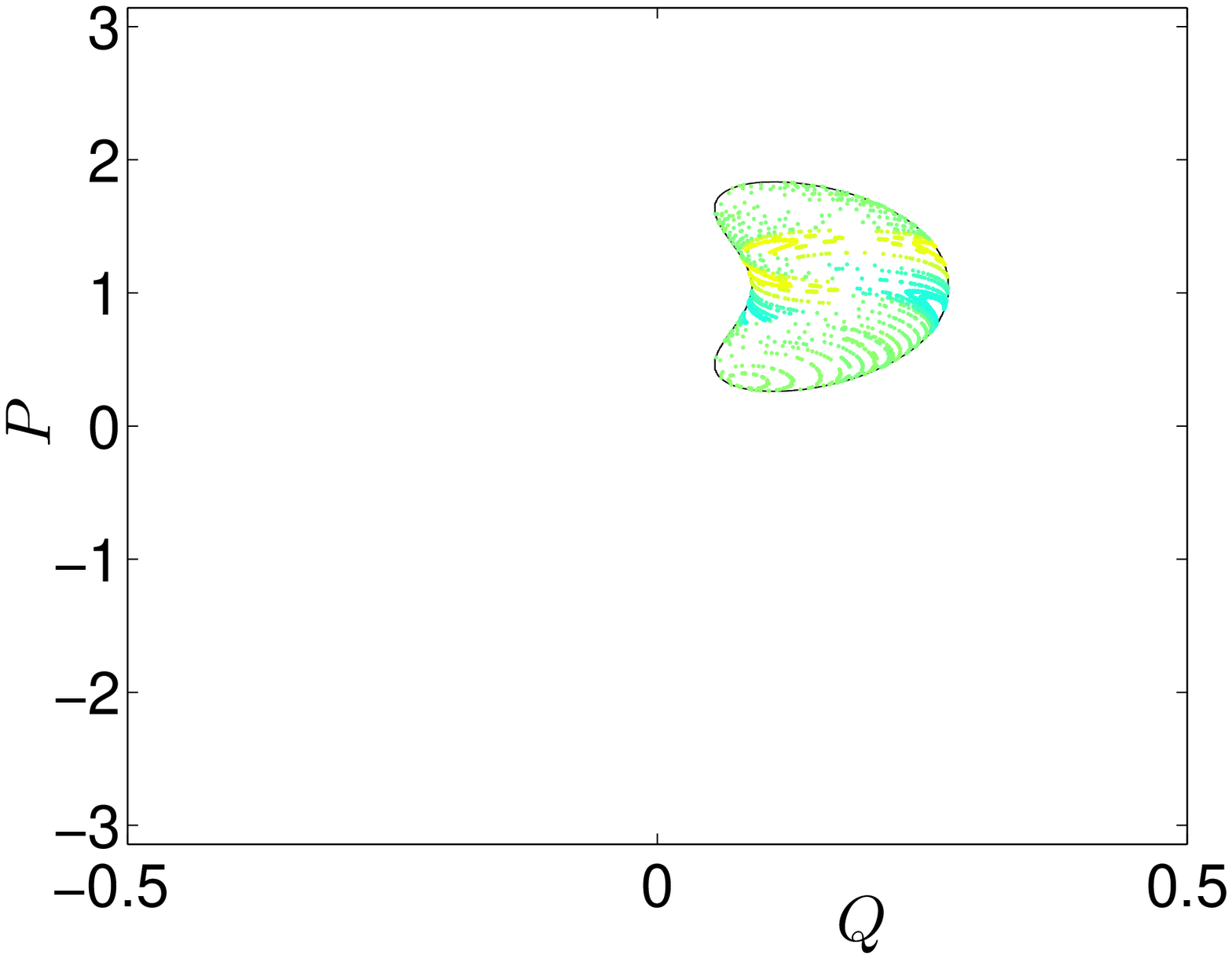} 
\includegraphics[width=0.24\hsize]{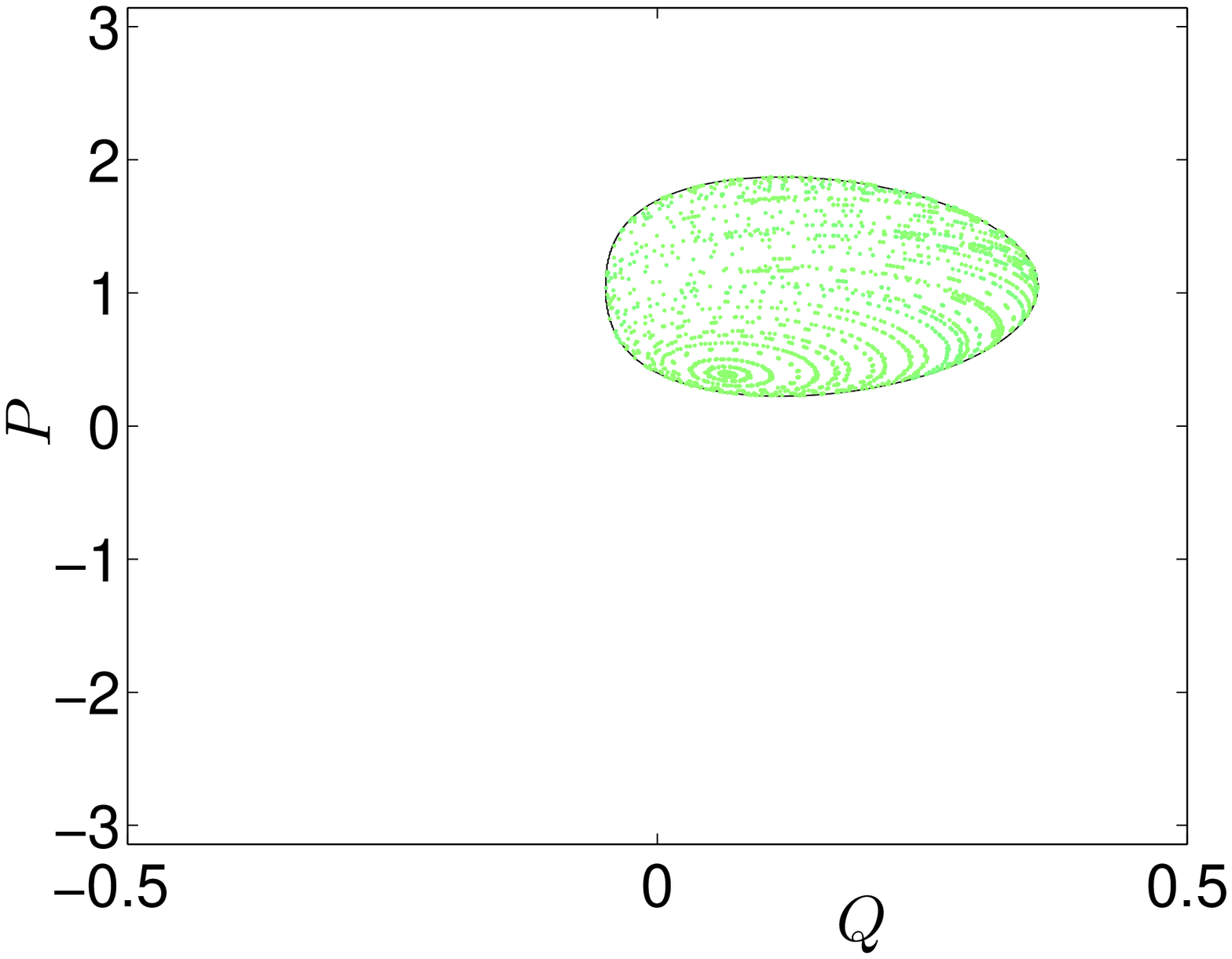}

\caption{ \label{fig2}  
Quantum spectrum (upper panels) and phase-space landscape (lower panels). 
The quantum spectra are for an $M{=}3$ ring with $N{=}45$ particles, 
dimensionless interaction ${u=2.5}$,
and weak-link coupling ratio ${K^{\prime}/K = 1,0.8,0.65,0.4}$ (from left to right). 
Axes and and color code are the same as in \Fig{fg_bloch1}.  
In each case an ${n_3 {-} n_1 =0}$ Poincare section is displayed.
The section coordinates are ${Q=(n_1-n_2)/(2N)}$ and ${P=(\varphi_1-\varphi_2)}$. 
The energy is chosen to be slightly above the ground state. 
The solid black line marks the borders of the allowed phase-space regions.
The outer regions are ``forbidden" energetically.  
The color code represents the averaged current for each classical trajectory:
red for larger clockwise current; blue for large anti-clockwise current; 
and yellow-to-green for very small current.
} 
\end{center}
\end{figure}

In this section we discuss what happens if a weak-link is introduced
into a ring that has a small number of sites ($M=3,4$).
In particular we ask what remains of the JCH phenomenology.
The first implication of the JCH phenomenology is the prediction of 
a critical $\alpha$ below which a quasi-degenerate doublet of flow 
states cannot exist.  If we naively use \Eq{e9} we deduce 
that the condition ${\alpha>1}$ for getting such doublet 
is $K^{\prime}/K > 1/2$ for $M=3$ and $K^{\prime}/K > 1/3$ for $M=4$. 
In order to inquire what is the actual threshold 
we plot quantum spectra for various values of $u$ and $K^{\prime}/K$. 
See \Fig{fig1}. We look for doublets at the bottom of the spectrum. 
A practical measure for that is $\mathcal{M}=  [\trc(\rho^2)]^{-1}$, 
where $\rho$ is the reduced one-body probability matrix, see \ref{a:fm}. 
The value of $\mathcal{M}$ indicates the {\em fragmentation} 
of the many-body state. It is $\mathcal{M}=1$ for a coherent state, 
and $\mathcal{M}\sim M$ for a quantum-ergodic state. 
In the case of a doublet the ground-state becomes 
a superposition of two coherent states hence $\mathcal{M}\sim 2$.  
Looking at \Fig{fig1} we see that for rings with $M=3,4$ sites, 
the $\alpha$ border is slightly higher than expected. 
We have verified using Poincare sections (see below) 
that for large $u$ the border is in agreement with ${\alpha_c=1}$.
For completeness we also show that for very large~$u$ (of order $N^2$) 
the value of $\mathcal{M}$ for the ground-state becomes of order~$M$, 
reflecting the Mott transition~\cite{sfs}.

To understand what determines the $\alpha_c$ border 
we display in \Fig{fig2} so-called Poincare sections 
of classical trajectories that are generated 
by the Hamiltonian \Eq{e2}. Namely, for display purpose 
a pair of canonical coordinates ${(Q,P)}$ is selected, 
and for each trajectory the sequence of points ${(Q(t_j),P(t_j))}$ 
where it intersects a specified phase-space section is recorded.
We see clearly that in the $\alpha<\alpha_c$ regime the two stability islands 
merge, reflecting that we no longer have the ``double well" 
structure in phase space.  

\begin{figure}
\begin{center}

\includegraphics[width=6cm]{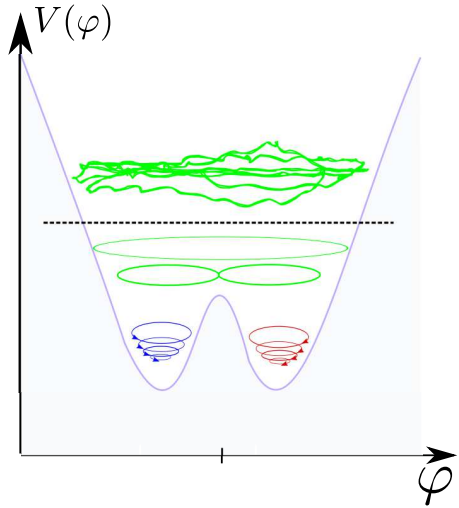}

\caption{ \label{fig2a}  
The energy landscape of the Josephson circuit Hamiltonian. 
Here the vertical axis represents the energy $E_{\varphi}$ 
of the weak-link DOF (the total energy $E$ should include the bath DOFs as well).    
The dashed line indicates the threshold $E_u$ for chaotic 
motion. Trajectories below $E_u$ are quasi-regular.
The JCH description is valid if $E_u$ is located well above $E_b$.   
} 
\end{center}
\end{figure}

\begin{figure}
\begin{center}

\includegraphics[width=4cm]{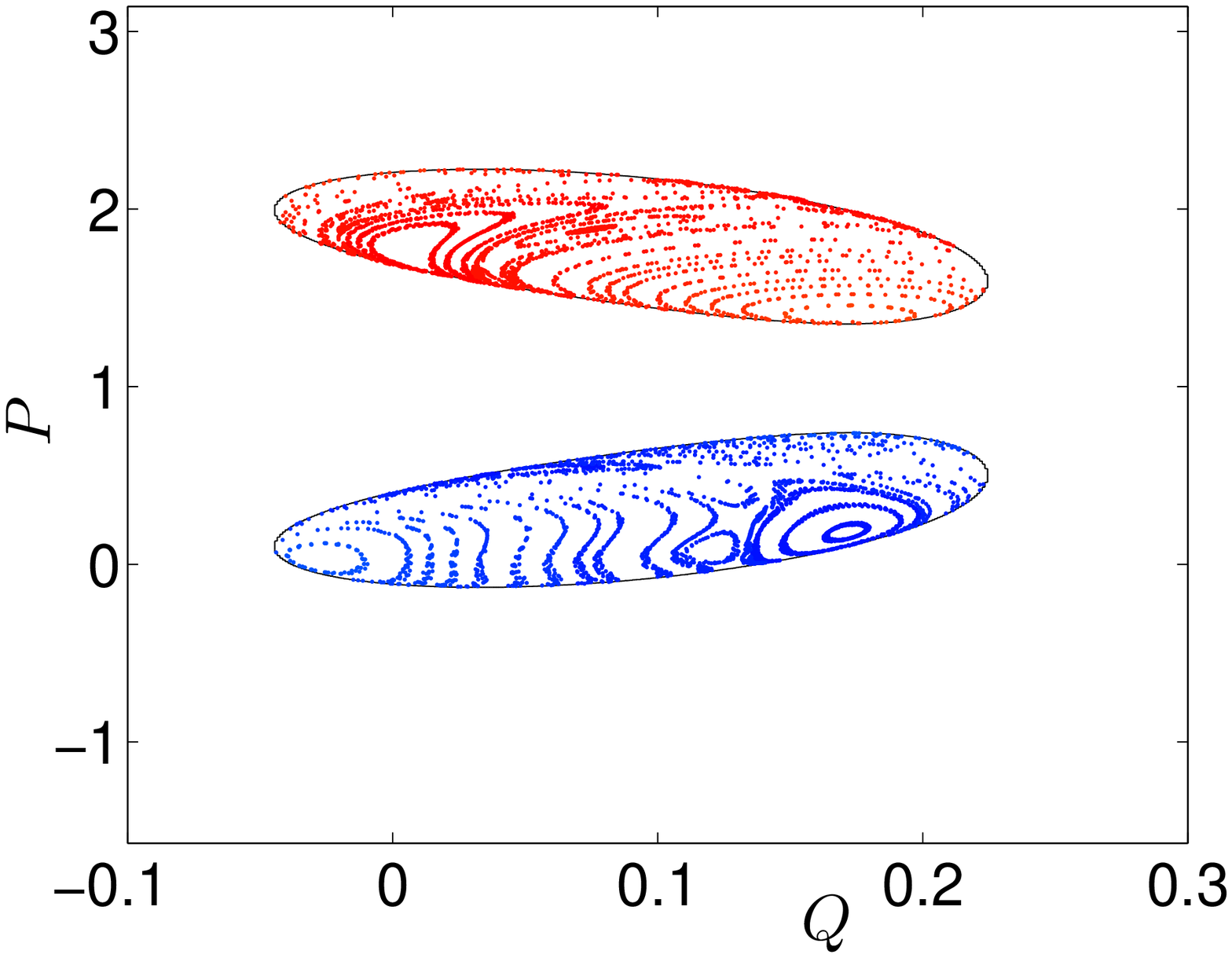} 
\includegraphics[width=4cm]{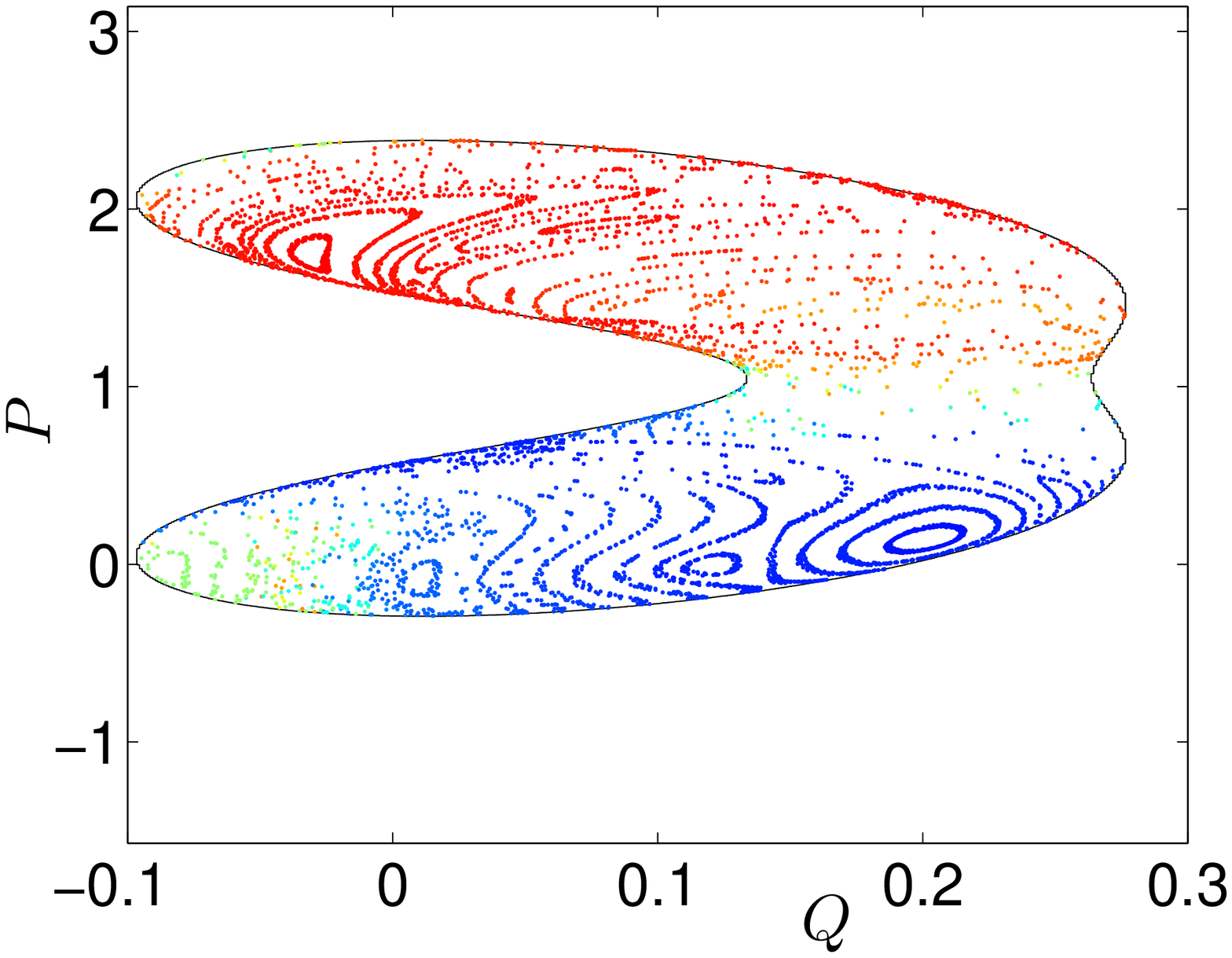} 
\includegraphics[width=4cm]{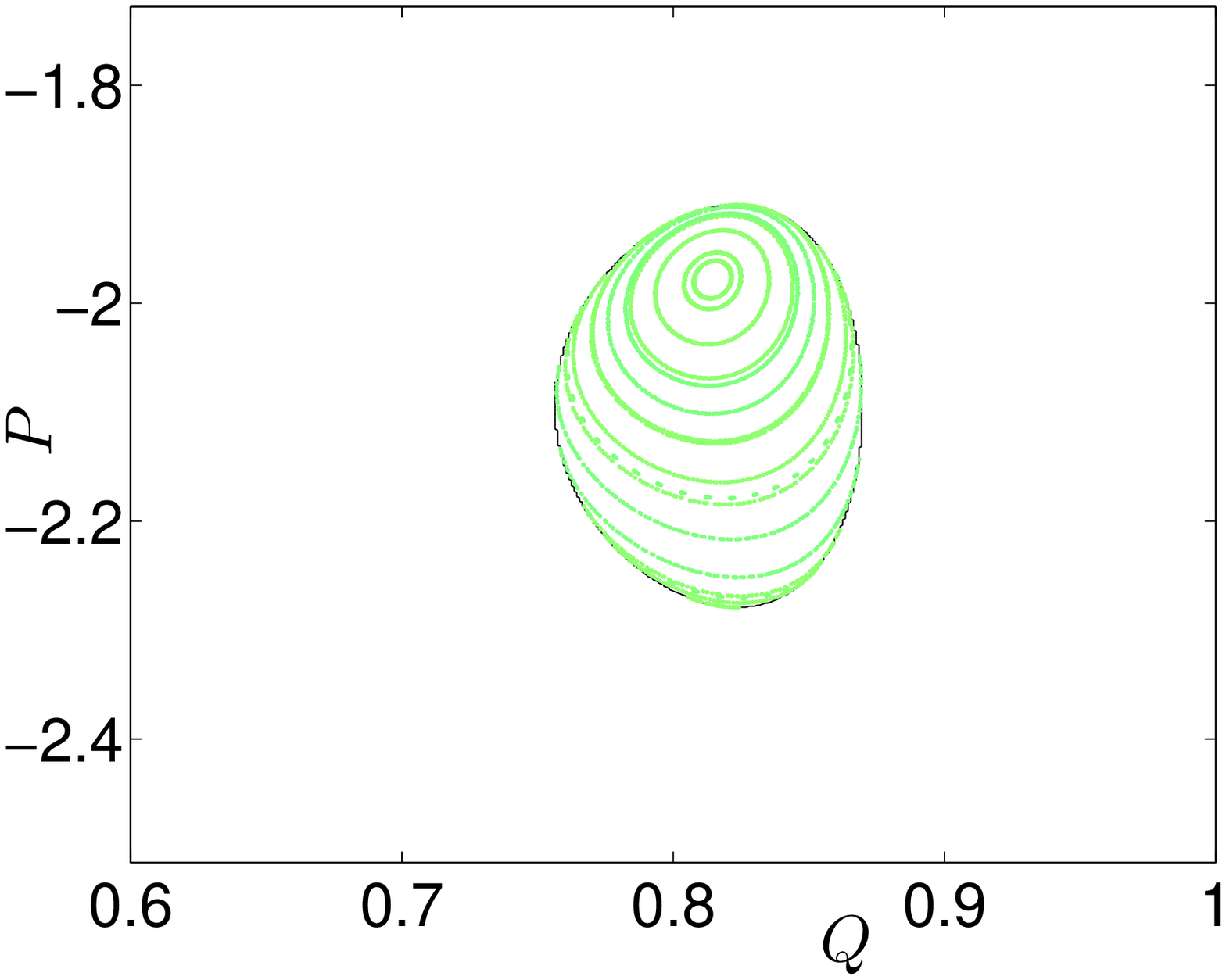} 

\caption{ \label{fig2b}  
Poincare sections at different energies. 
Panel (a) is a zoomed version of \Fig{fig2}b. 
Panels (b) and (c) are for the same model parameters 
but the total energy is, respectively, 
${E/N=-0.036}$ (slightly above the barrier energy $E_b$) 
and ${E/N=1.48}$ (close to the upper most energy in the spectrum). 
In each panel all  the trajectories have the same total energy. 
But if we subtract the bath energy, they correspond to 
the different trajectories of \Fig{fig2a}. 
In panel~(c) the island contain self-trapped trajectories, 
hence it can support self-trapped states (condensation in one site). 
This should be contrasted with panel~(a) where the two islands  
can support different flow-states (condensation in momentum).    
} 

\ \\

\includegraphics[width=4cm]{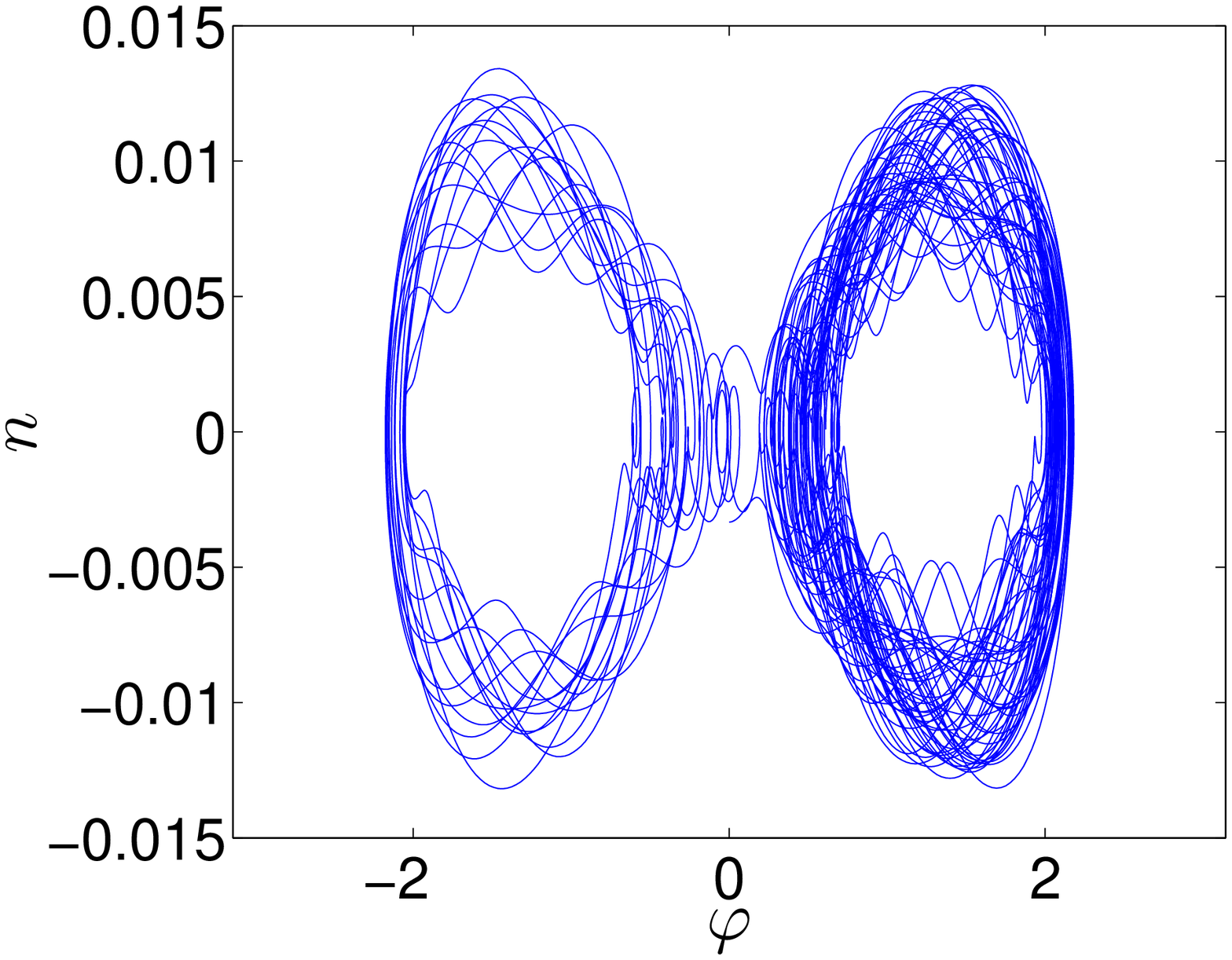} 
\includegraphics[width=4cm]{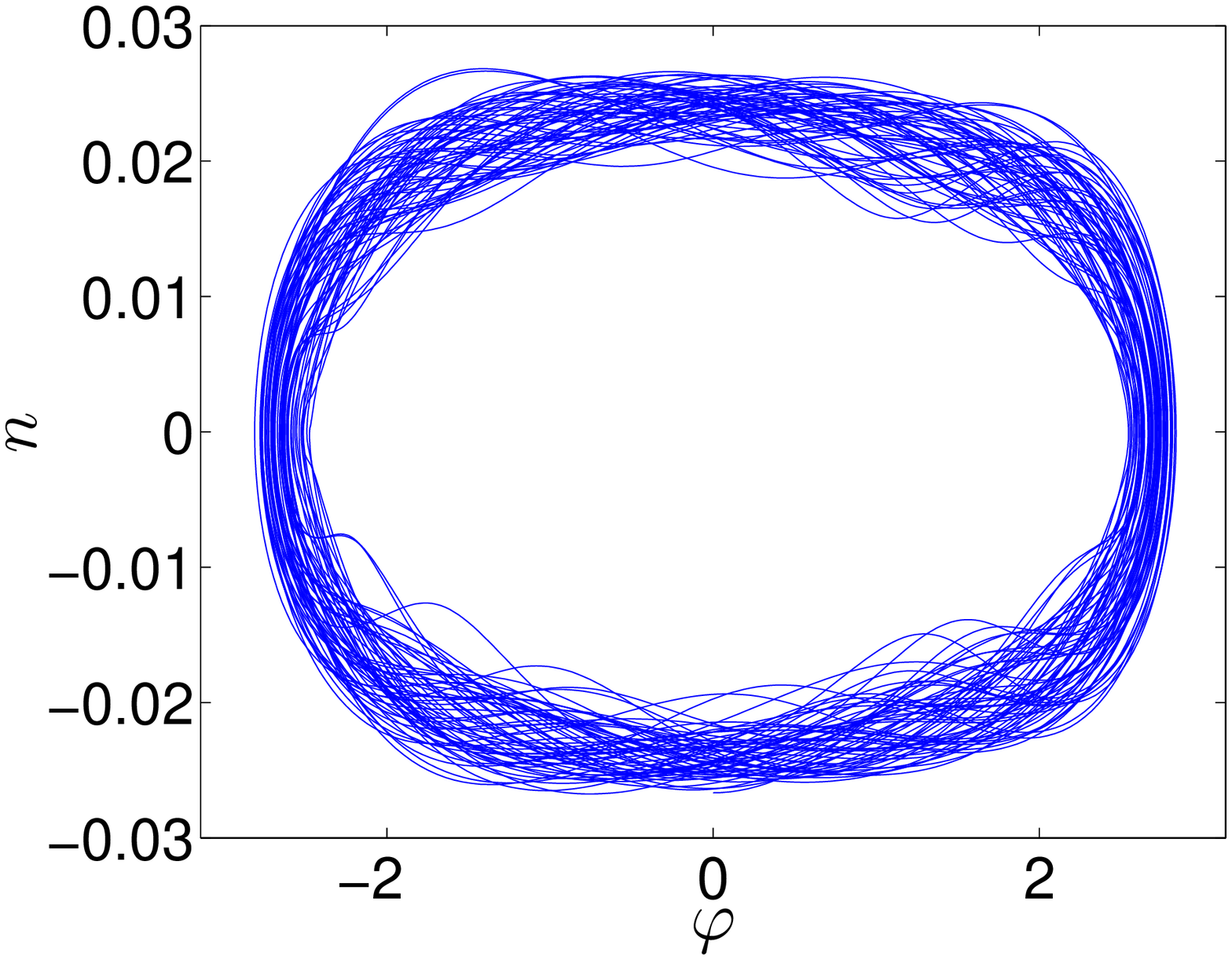}
\includegraphics[width=4cm]{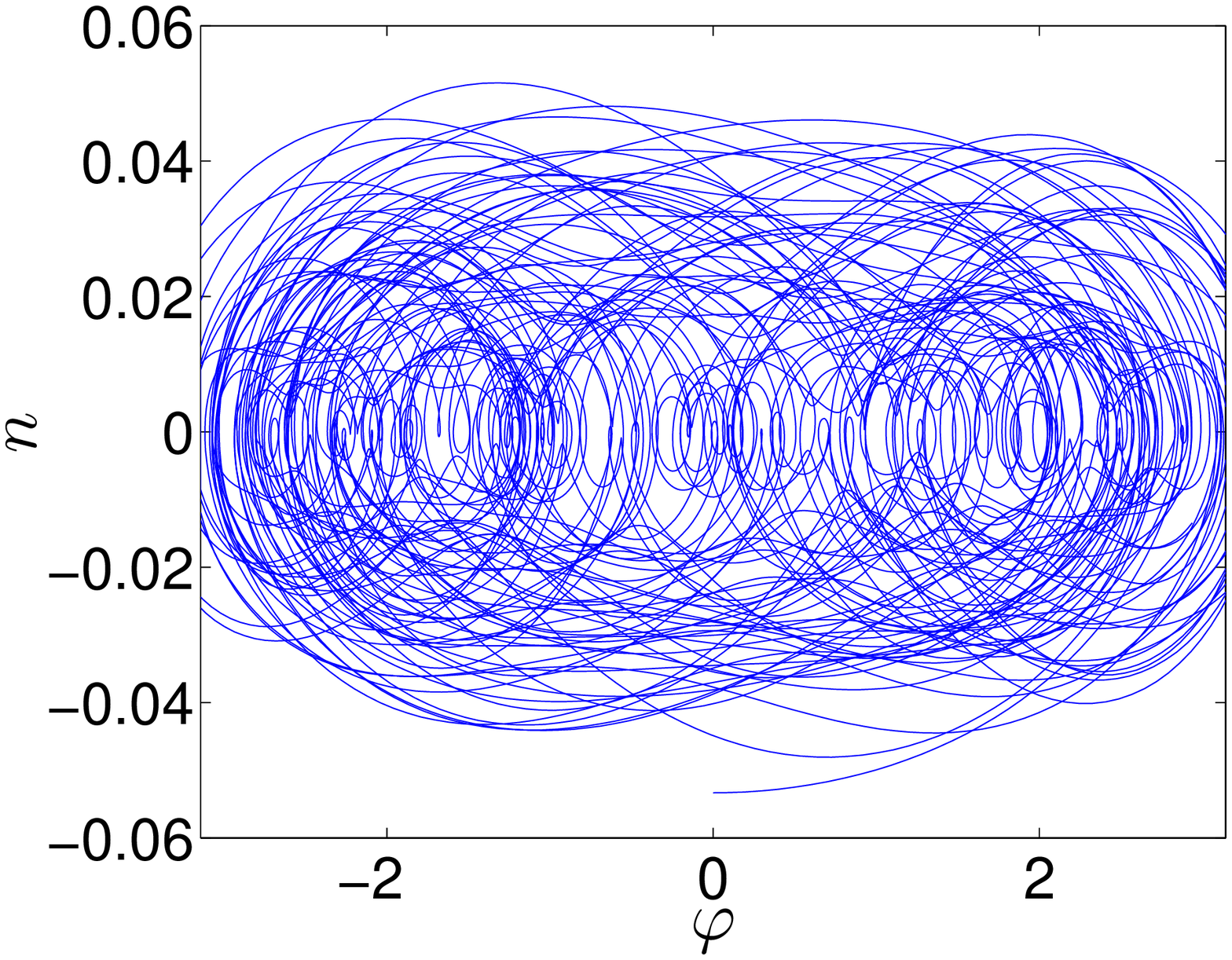}

\includegraphics[width=4cm]{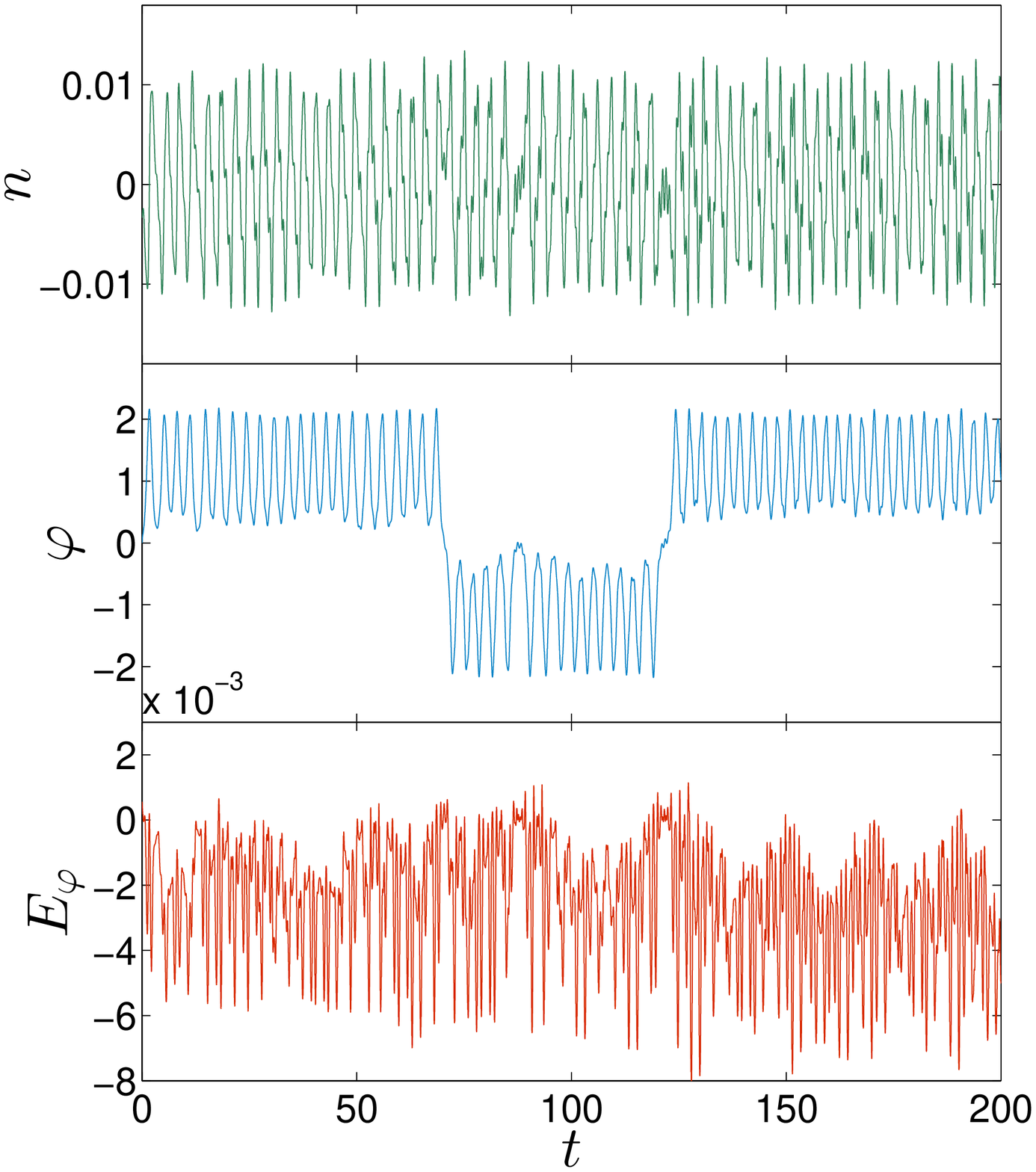} 
\includegraphics[width=4cm]{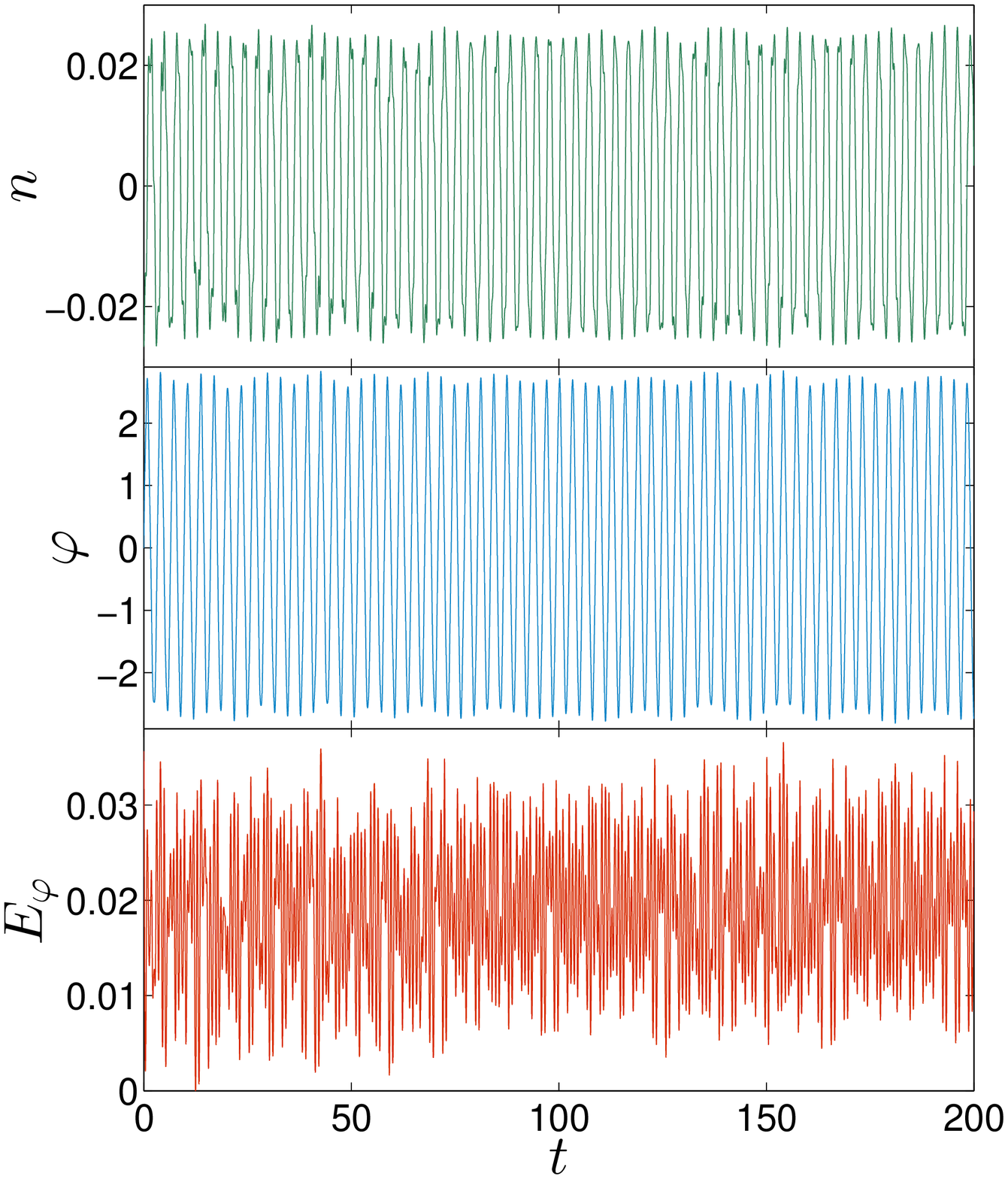}
\includegraphics[width=4cm]{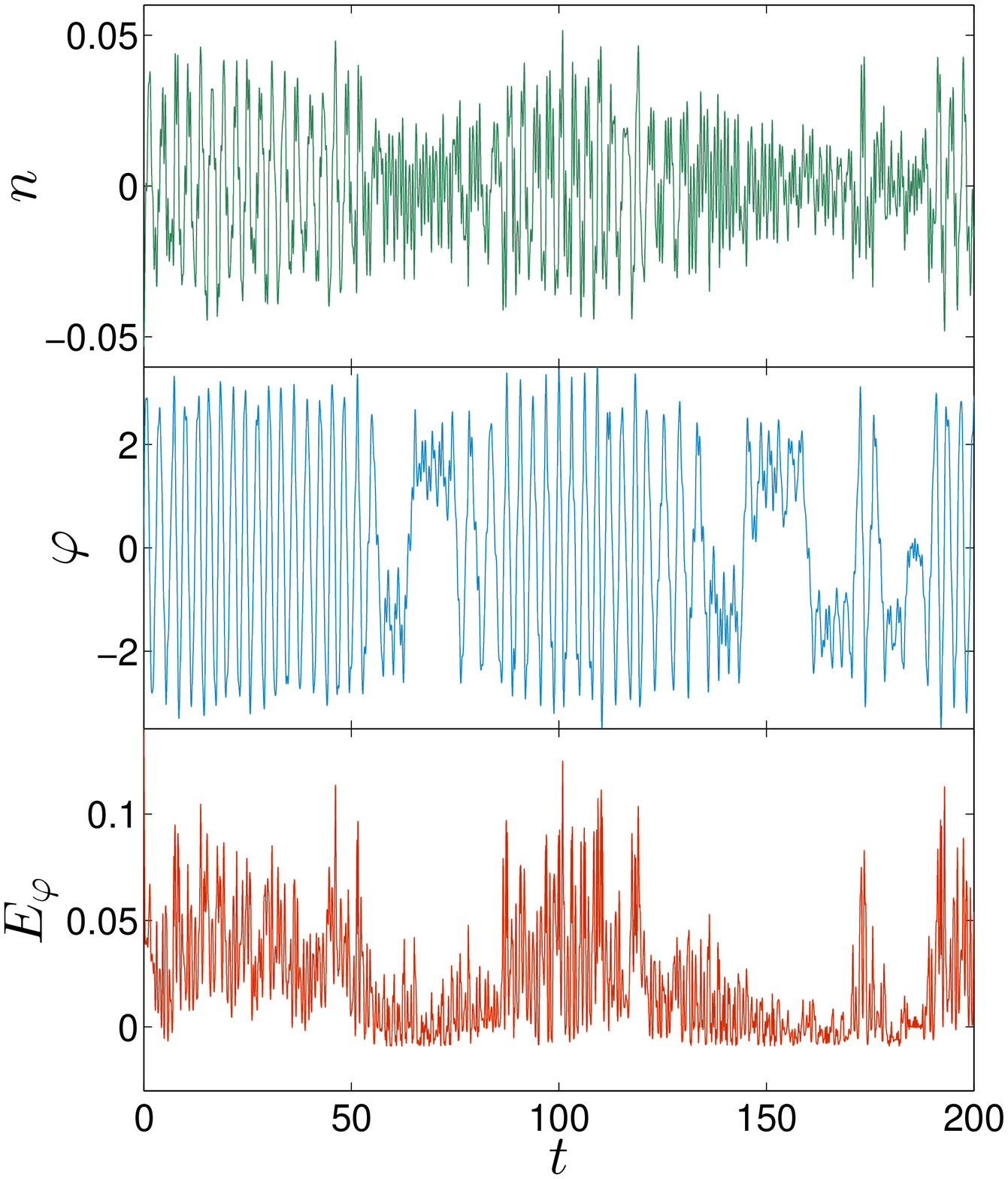}

\caption{ \label{fig3}  
Representative $(\varphi(t),n(t))$ trajectories of an $M{=}6$ ring with weak-link.
The lower panels are for the weak-link energy $E_{\varphi}(t)$ measured relative to the top of the barrier.
The system parameters are ${u{=}200}$, and ${K_J/K{=}0.3}$, and ${\Phi{=}\pi}$. 
In all the panels the initial condition is in the vicinity of the barrier, 
with equal populations $n_j=N/6$. 
The actual starting point is with ${n_1=(1/6+\delta)N}$ and ${n_6=(1/6-\delta)N}$.    
In (a) the junction energy is mostly below the barrier (${\delta=0.01}$),  
and we see that the dynamics is in qualitative agreement with the JCH: 
we observe regular flow-motion with rare jumps 
to the opposing flow-motion due to an activation by the ``bath" DOFs. 
In (b) the junction energy is above the barrier (${\delta=0.08}$), 
and we still observe pendulum-like regular motion.
In (c) the energy is above the chaos threshold (${\delta=0.16}$), 
and we get irregular chaotic motion that is no longer described by the JCH.
This should be contrasted with the $M=3$ trajectories 
of \Fig{fig2b}b where the chaos threshold $E_u$ 
coincides with $E_b$, invalidating the JCH phenomenology.    
} 
\end{center}
\end{figure}

But this is not enough. The JCH should be trusted 
also when we analyze tunneling or phase-slips through 
the {\em forbidden region}. For this purpose 
it should describe correctly the dynamics up to 
some energy well above the barrier. 
This means the the threshold $E_u$ for chaotic motion 
should be above the threshold $E_b$ for barrier crossing.
See illustration in \Fig{fig2a}.
We therefore plot in \Fig{fig2b}b, a Poincare section
for an energy that is slightly above $E_b$.
What we see is that trajectories that go across 
the barriers are chaotic rather than regular. 
This indicates that a JCH description of the dynamics 
is in fact {\em not} valid.  

Let us try to understand the reason for the failure 
of the JCH description. In the vicinity of a single flow-state 
worst case scenario is that a phase difference $\pi$ 
has to be supported by the ring. The harmonic 
approximation requires $\pi/(M-1)<\pi/2$ on each bond.
This is marginally satisfied for $M=3$. But if we want 
the JCH description to be valid over a $2\pi$ range of $\varphi$, 
then the requirement becomes  $2\pi/(M-1)<\pi/2$,    
meaning we have to consider rings with ${M \ge 5}$ sites.
Similar claim has appeared in \cite{Hekking}. 
In \Fig{fig3} we verify that for an ${M=6}$ ring with 
weak-link the chaos border $E_u$ is indeed well above the 
barrier energy~$E_b$. Up to $E_u$ the dynamics looks 
like that of a pendulum that is slightly affected by the 
other ``bath" DOFs. Above $E_u$ the motion becomes chaotic 
and the JCH description is no longer applicable.

\section{weak-link in a many site ring}

Consider $N$ bosons is ring of length $L$, such that 
the average density is $\rho=N/L$. The so called Lieb-Liniger parameter 
that controls the quantum aspect of the interaction is ${\gamma= \mass g / \rho}$.
For ${\gamma \gg 1}$ the hard-core bosons are like fermions, 
while for  ${\gamma \ll 1}$ we can use a ``classical" description.
In the latter case the ``trajectories" obey the so-called Gross-Pitaevskii (GP) equation.   
In fact the parameter $\gamma$ does not appear in the GP treatment 
of the model. The only dimensionless parameter of the GP description is 
\be{17}
u_L \ \ = \ \ N^2\gamma \ \ =  \ \ N L \mass g
\eeq 
We shall refer to it as the ``classical" 
dimensionless parameter, while ${\hbar=1/N}$ can be  
regarded as the dimensionless Planck constant. 
Within the framework of the ``classical" (GP) treatment 
the low excitations of the systems are phonons 
with sound velocity $c=(g\rho/\mass)^{1/2}$.
For a finite length ring the spacing in the frequencies 
of the phononic modes is ${\Delta_0 = \pi c / L}$.

If we add a periodic potential that divides the ring into $M$ sites, 
we get a system that possibly can be described by the BHH \Eq{e1}.
The analogue of the GP is the discrete nonlinear Schrodinger (DNLS) equation.
The distance between the sites is $a=L/M$ and the average number 
of particles per site is ${\bar{n}=N/M}$. The effective parameters 
of the BHH are accordingly $U=g/a$ and 
\beq
K \ \ = \ \ \frac{1}{\mass a^2}\eexp{-S_0} \ \ \equiv \ \ \frac{1}{\mass^* a^2}
\eeq     
where $S_0$ reflects the height of the barrier.
The effective quantum parameter is 
\beq
\gamma^* \ \ \equiv \ \ \frac{\mass^* g}{\rho} \ \ = \ \ \gamma \eexp{S_0} \ \ = \ \ \frac{U}{\bar{n}K}   
\eeq
This parameter controls the quantum Mott transition. 
Namely for ${\gamma^* >1}$ superfluidity is diminished 
if ${\bar{n}}$ is close to integer.   
In addition we can define the ``classical" dimensionless parameter 
which is analogous to $u_L$ of \Eq{e17} as  
\beq
u_M \ \ = \ \  M u \ \ = \ \ N^2 \gamma^* 
\eeq
The $u_M$ parameter controls the DNLS equation, and determines the stability 
of the steady flow solutions, as well as the thresholds for self-trapping 
and soliton formation. Due to the discretization we have effectively $M$ phononic modes, 
whose spectrum is charaterized by the cutoff frequency 
\beq
\omega_c \ \ = \ \  \left( \bar{n}U K \right)^{1/2} \ \ \sim \ \ M \ \Delta_0
\eeq   
Where $\Delta_0$ is formally the same as for a continuous ring, but with $m^{*}$.  

For a regular ring with a weak-link the reduction to an effective JCH provides   
the following expressions \cite{Dimitry}:  
${E_C=g/L}$ and ${E_L=\rho/(\mass L)}$ and ${E_J = \alpha E_L}$. 
The parameter $\alpha$ is controlled by the tunnel-coupling,  
which is determined by the height of the barrier at the weak-link.   
Our derivation in \ref{a:jch} has provided similar 
expressions, but there are some differences.
First of all the effective mass is of course $\mass^*$ and not $\mass$, 
and therefore the effective quantum parameter $\gamma$ becomes  $\gamma^*$.
A secondary difference is that $E_C=g/L$ is replaced by $E_C=g/L^*$, 
where the effective length over which the density varies 
might be as small as ${L^*=a}$. The latter value reflects the extreme case 
of uniform distribution of the particles along the ring.  
Expression \Eq{e9} for the parameter $\alpha$ can be written as   
\beq
\alpha \ \ \equiv \ \ \frac{E_J}{E_L}
\ \ = \ \ M \ \eexp{-(S_J-S_0)}
\eeq
where $S_0$ and $S_J$ reflect the heights of the barriers
in regular bonds, and at the weak-link respectively.

We turn our attention to the bath. 
The derivation in \ref{a:jch} shows that
within the bilinear-coupling approximation the effective 
number of bath DOFs is $d_{\text{bath}}=\lfloor (M{-}2)/2 \rfloor$.
Consequently the bath Hamiltonian 
has the familiar Caldeira-Leggett form \Eq{e3}, 
with $\mass_m=1/U$, 
and $\omega_m^2 = 2 U K \bar{n} (1- \cos k_m)$, 
and $ k_m = \pi m / (M-1)$, 
and $c_m=K\bar{n}[2/(M-1)]^{1/2} \sin k_m$. 
From that follows that the dissipation coefficient is 
\beq
\eta \ \ = \ \ \frac{\pi}{\sqrt{\gamma^*}}
\eeq
In Ref.\cite{Dimitry}, regarding regular ring, it has been claimed 
that if $(E_J/N) \ll \Delta_0$ (called there ``the small ring limit") 
then  the bath can be ignored. 
In the context of the present Bose-Hubbard circuit this condition takes 
the form $K^{\prime} \ll \omega_c$, 
meaning that the bath should have high frequency 
cutoff compared with the hopping rate.
But from the work on the spin-boson problem 
we know that the condition for witnessing coherent 
oscillation is $\eta<\pi$ which implies  
that $\gamma^*$ should be large compered with unity.
We identify that this is a problematic non-semiclassical regime 
where the Mott transition takes place. 
Namely, for ${\gamma^*>1}$ the superfluidity of the system 
depends sensitively on the filling ratio~$N/M$. 
In a grand-canonical perspective the system has the tendency 
to become a Mott insulator.

\clearpage
\section{Discussion} 

We observe that a TLS modeling of quasi-degenerate flow-states 
in a few-site ring is feasible, meaning that coherent Rabi oscillations 
are not over-damped. This is true with or without a weak-link, 
and the frequency is possibly determined by chaos-assistance tunneling. 
In particular we have demonstrated numerically Rabi oscillations 
between metastable flow-states in a non-rotating ($\Phi=0$) 
circuit that consists of $M=4$ sites.

We have determines what is the minimal value of $\alpha$ 
that does not endanger the meta-stability of the $\Phi=\pi$ flow-states.
Clearly below this minimal value a weak-link is not useful.
From a semi-classical perspective this value is the threshold 
for the merging of two stability islands. 
For large rings, assuming that the JCH phenomenology is valid, 
the minimal value is implied by the familiar condition ${\alpha>\alpha_c}$
with ${\alpha_c=1}$
We note that in a super-conducting circuit, due to the Meisner effect,  
the effective inductance is larger, and $\alpha$ is typically large. 

In the semiclassical perspective the flow-states are supported by 
a local minimum of the energy landscape (Landau stability), 
or by a region that is surrounded by KAM tori. 
In the latter case, for rings with ${M>3}$ sites the stabilization is due to a many-body quantum 
localization effect, that suppresses the Arnold diffusion. 
Depending on the type of states involved, the coupling 
might be via a forbidden-region (as implied by the JCH phenomenology), 
or it might be mediated by a chaotic sea. In the latter case 
the chaos-assisted tunneling provides a weaker dependence on the number 
of particles involved.   


{\em The system plus bath perspective.--} 
Formally the circuit has ${d=M{-}1}$ interacting DOFs, 
while in the approximated JCH version we have a single DOF $(\bm{\varphi},\bm{n})$ 
that interacts with a ``bath" that consists of a few DOFs. 
If the bath is ignored the motion in the single DOF phase-space 
is regular, and looks formally the same as that of a pendulum. 
If ${\alpha>\alpha_c}$, a separatrix is formed, 
hence we have two stability-islands that can support the two 
quasi-degenerate flow-states. But if the bath is taken into account, the projected 
motion in the  $(\bm{\varphi},\bm{n})$ coordinates becomes ``dressed" and ``noisy",  
in the same sense as discussed by Caldeira, Leggett and followers.
These effects endanger the coherent Rabi oscillations.

{\em Large $M$ ring.-- }
For a regular ring with bosons one can define the Lieb-Liniger parameter $\gamma$. 
Having $\gamma>1$ means that quantum effects become important (GP description becomes problematic), 
but nevertheless there is no quantum phase-transition. For the BHH ring (bose gas in an optical lattice), 
we have defined an effective $\gamma^*$ that corresponds to the effective mass in the lattice. 
As before $\gamma^*>1$ means that quantum effects are important. But here the consequences are much more 
dramatic. Namely, the quantum regime $\gamma^*>1$ is identified as the Mott-regime, 
where depending on the filling-ratio the ring can become a Mott-insulator.
On the other hand the analysis shows that $\gamma^*>1$ is the condition for witnessing coherent Rabi oscillations.  
So there is clash here: on the one hand we want the ring to be in a superfluid phase (avoid Mott); on the other 
hand we want to have weak coupling to the bath in order to witness coherent oscillations.
Thus for a many-site ring  the requirement for observing coherent Rabi oscillation might be in clash with the quantum Mott transition.

{\em Small $M$ ring.-- }
We wanted to understand how this standard JCH phenomenology is modified 
if the ring consists of a small number of sites. Then the ``bath" consists of 
a small number of DOFs and the standard Caldeira-Leggett perspective 
becomes questionable. One direction \cite{dfs}
is to say that the interaction with 
{\em chaotic} DOFs is essentially like the interaction with infinitely many 
{\em harmonic} DOFs, hence coming back to Caldeira-Leggett phenomenology. 
This type of argument might work for rings with $M \ge 6$ sites 
for which the effective number of bath DOFs is ${d_{\text{bath}} \ge 2}$. 
We did not take this route here. Rather we discussed the whole issue in a much 
more fundamental level, focusing on rings with ${M=3,4,5}$ sites.

Arnold diffusion is in a sense the low dimensional version of having a ``bath". 
The essence of Arnold diffusion is that a selected DOF does not perform 
an unperturbed integrable (pendulum like) motion. Rather the motion 
always ``diffuses" due to the ``noise" that is induced by the other DOFs. 
Hence we have here a formal equivalence with the ``system plus bath" perspective. 
It follows rigorously that a necessary condition for the applicability 
of the ``system plus bath" paradigm with regard to a circuit with a weak-link 
requires more than 3~sites. But this is not a sufficient condition. 
We have emphasized that a JCH modeling implies regular motion up to 
an energy that exceeds the barrier height. Such high threshold for chaos 
is apparently feasible only for rings that have more than 5~sites. 

\ \\

\noindent
{\bf Acknowledgements.-- }
We thank Luigi Amico for motivating the present study.
This research has been supported by  by the Israel Science Foundation (grant No. 29/11).

\appendix
\clearpage

\section{Superfluidity in low dimensional circuits}
\label{a:sf}

In this Appendix we provide a brief summary for the ``big picture" of mesoscopic superfluidity.
The key issue is the meta-stability of the flow-states.
We follow \cite{sfc}, while some preliminaries regarding the energy landscape 
and the dynamical stability issues can be found in \cite{sfs} and \cite{niu} respectively. 

In the conventional ``Landau criterion" picture the flow-states 
are energetically stable, i.e. they are located in local 
minima of the energy landscape. Hence they are separated  
by a ``forbidden region" and the coupling requires tunneling.

But metastability can be achieved even in the absence 
of energetic-stability. For $M=3$ ring, the flow-state 
can be dynamically stable, protected in phase-space by Kolmogorov Arnold and Moser (KAM) tori. 
Then the generic picture is two islands that are separated 
by a chaotic sea, and not by a forbidden region. 

For $M>3$ rings, the KAM tori are not able to divided 
phase-space into territories. The dynamics takes place 
on an ``Arnold web" of resonances. This leads to so-called 
Arnold diffusion: if we look on the weak-link degree of 
freedom ${(\varphi,n)}$ we expect to see diffusion of its energy.
We emphasize that such diffusion does not occur 
in $M=3$ ring: there it is arrested by the KAM tori.  

The discussion above might give the impression 
that flow-states cannot survive in $M>3$ rings. 
But in fact quantum mechanics saves us: dynamical stability 
can be maintained in-spite of Arnold diffusion. 
This can be regarded as a many-body localization effect.
It follows from the following simple consideration: 
The time to escape an Arnold web region might be very long; 
if the required time is larger than the quantum breaktime 
(inverse level spacing) then the escape will never happen.

\section{Definition of the fragmentation measure $\mathcal{M}$}
\label{a:fm}

The eigenstates of the Hamiltonian \Eq{e1} 
can be characterized by their fragmentation ${\mathcal{M}= [\trc(\rho^2)]^{-1}}$,
where the one-body reduced probability matrix is 
\beq
\rho_{ij} \ \ = \ \ \frac{1}{N} \, \langle \bm{a}_j^{\dag} \bm{a}_i\rangle
\eeq
Roughly speaking $\mathcal{M}$ tells us how many orbitals are occupied by the bosons. 
A value of ${\mathcal{M}=1}$ indicates that the state it not fragmented, 
hence it can be written as $ ( b^{\dagger}_k )^N | \text{vacuum} \rangle $. 
Here $b^{\dagger}_k = \sum_j c_j^k a^{\dagger}_j $ creates a particle in some superposition of the site modes, 
with coefficients $c_j^k$. Such states are the many-body coherent-states
in the generalized sense of Perelomov \cite{perelomov}. 
Their phase-space representations are minimal wave packets situated at some
point $(\bm{\varphi},\bm{n})$ of phase space.
A higher value ${1 < \mathcal{M} \leq M}$ indicates that 
the bosons are fragmented into several orbitals.

\clearpage
\section{Derivation of the Josephson Circuit Hamiltonian}
\label{a:jch}

Consider $N$ Bosons in an $M$~site ring described by the BHH \Eq{e1}.
In the limit $u \gg M$ it is common to neglect the fluctuations of the number of atoms 
in each well \cite{Paraoanu}, and approximate the Bose-Hubbard model 
with the so called quantum-phase-model (``coupled rotors")
which is formally equivalent to an array of Josephson junctions:
\beq
\mathcal{H} \ \ = \ \  \sum_{j=1}^{M} 
\left[
\frac{U}{2} \bm{n}_{j}^2 
- \bar{n} K_j  \cos\left((\varphi_{j{+}1}{-}\varphi_{j}) - \frac{\Phi}{M}\right) 
\right]   
\eeq
Where $\bm{n}_j $ and $\varphi_j$ are canonically conjugate variables. Without lost of generality, we can employ a gauge transformation such that the phase $\Phi$ vanishes at all bonds except the weak-link. Namely,
\beq
\mathcal{H} \ = \ \sum_{j=1}^{M}
\frac{U}{2} \bm{n}_{j}^2 
-\bar{n}K  \sum_{j=1}^{M-1}  \cos\left(\varphi_{j{+}1}{-}\varphi_{j} \right)
-\bar{n}K^{\prime} \cos\left(\varphi_1 {-}\varphi_{M} -\Phi\right)
\eeq
With a weak-link $K^{\prime} \ll K$, the phase difference at the $M-1$ regular bonds becomes small such that $\cos\left(\varphi_{j{+}1}{-}\varphi_{j} \right)\sim 1 $. The Hamiltonian can then be written, up to a constant, as:
\beq
\mathcal{H} \ = \  \sum_{j=1}^{M}
\frac{U}{2} \bm{n}_{j}^2 
+\frac{\bar{n}K}{2} \sum_{j=1}^{M-1} (\varphi_{j{+}1}{-}\varphi_{j})^2 
-\bar{n}K^{\prime} \cos\left(\varphi_1 {-}\varphi_{M} -\Phi\right) 
\eeq
The second sum can be written as:
\beq
&& \sum_{j=1}^{M-1} (\varphi_{j{+}1}{-}\varphi_{j})^2 \ = \ 
\varphi_1^2 + \varphi_M^2 -2 \varphi_1 \varphi_2 -2 \varphi_{M{-}1} \varphi_M  + \sum_{i,j=2}^{M-1} A_{ij} \varphi_i \varphi_j   
\\
&& = \ \frac{\varphi_+^2}{2} + \frac{\varphi_-^2}{2} - \varphi_+ ( \varphi_2 + \varphi_{M-1}) - \varphi_- ( \varphi_2 - \varphi_{M-1}) + \sum_{i,j=2}^{M-1} A_{ij} \varphi_i \varphi_j  
\eeq
Where we introduced the notation $\varphi_\pm = \varphi_1 \pm \varphi_M$, and $A_{ij} = 2 \delta_{ij} -\delta_{i,j\pm 1}$. 
Consequently 
\beq
\mathcal{H} \ &=& \
\frac{U}{4} \left( \bm{n}_{-}^2 +  \bm{n}_{+}^2 \right)  
+ \frac{\bar{n}K}{4} \left( \varphi_-^2  + \varphi_+^2 \right) 
\\ \nonumber 
&& - \bar{n}K^{\prime} \cos\left(\varphi_- -\Phi\right)  
-\frac{\bar{n}K}{2} \left[ \varphi_- ( \varphi_2 - \varphi_{M-1}) 
+ \varphi_+ ( \varphi_2 + \varphi_{M-1}) \right] 
\\ \nonumber
&& + \frac{U}{2} \sum_{j=2}^{M-1}	\bm{n}_{j}^2   
+ \frac{\bar{n}K}{2} \sum_{i,j=2}^{M-1} A_{ij} \varphi_i \varphi_j   
\eeq
The last line can be easily diagonalized:
\beq
\frac{U}{2} \sum_{j=2}^{M-1}	\bm{n}_{j}^2   +   \frac{\bar{n}K}{2} \sum_{i,j=2}^{M-1} A_{ij} \varphi_i \varphi_j   
\ \ = \ \
\sum_{m=1}^{M-2} \left( \frac{U}{2}	\tilde{\bm{n}}_{m}^2  +   \frac{\omega_m^2}{2U}  \tilde{\varphi}_m^2
\right)
\eeq
with 
\beq
\omega_m^2 &=& 2 U K \bar{n} (1- \cos k_m) \\
k_m &=& \pi m / (M-1) \\ 
\tilde{\varphi}_m &=& \sqrt{\frac{2}{M-1}} \times  \sum_{j=2}^{M-1}  \sin \left[ k_m (j-1) \right] \varphi_j
\eeq
Due to the reflection symmetry of the ``chain" ($j=2,..,M-1$), 
the $m=\text{odd}$ and $m=\text{even}$ modes are symmetric and anti-symmetric respectively.
The coupling term $\varphi_\pm ( \varphi_2 \pm \varphi_{M-1})$ can be expressed as follows: 
\beq
&& \varphi_\pm  \sqrt{\frac{2}{M-1}} \times  \sum_{m=1}^{M-2} \left[
\sin \left( k_m  \right) \pm \sin \left( k_m (M-2) \right) \right] \tilde{\varphi}_m 
\\ 
&& = \varphi_\pm  \sqrt{\frac{2}{M-1}} \times  \sum_{m=1}^{M-2} 
\sin \left( k_m  \right) \left[ 1 \pm (-1)^{m-1} \right] \tilde{\varphi}_m
\eeq
We see that $\varphi_{+}$ is coupled only to the symmetric modes ($m=\text{odd}$), 
while $\varphi_{-}$ is coupled only to the anti-symmetric modes ($m=\text{even}$).
With the above substitutions the Hamiltonian takes the form:
\beq
\mathcal{H} &=& 
\frac{U}{4} \left( \bm{n}_{-}^2 +  \bm{n}_{+}^2 \right)  + \frac{\bar{n}K}{4} \left( \varphi_-^2  + \varphi_+^2 \right) -\bar{n}K^{\prime} \cos\left(\varphi_- -\Phi\right)  \\
&-&  \varphi_- \sum_{m=even}^{M-2} c_m \tilde{\varphi}_m -\varphi_+ \sum_{m=odd}^{M-2} c_m \tilde{\varphi}_m
+ \sum_{m=1}^{M-2} \left( \frac{U}{2}	\tilde{\bm{n}}_{m}^2  +   \frac{\omega_m^2}{2U}  \tilde{\varphi}_m^2 \right) 
\eeq
with 
\beq
c_m \ \ = \ \ K\bar{n}[2/(M-1)]^{1/2} \sin k_m
\eeq 
The Hamiltonian consist of the two freedoms $\psi_\pm$ which are coupled to an harmonic bath of $M{-}2$ DOFs. 
But in-fact only the weak-link DOF $\psi_-$ and the $m=\text{even}$ modes of the bath are of interest. 
The freedom $\psi_+$ can be thought of as a part of the $m=\text{odd}$ modes of the bath, 
which does not interact with the weak-link. 
%
%
So that the relevant part of the Hamiltonian is:
\beq
\mathcal{H} &=& 
U \bm{n}^2  + \frac{\bar{n}K}{4} \varphi^2   -\bar{n}K^{\prime} \cos\left(\varphi -\Phi\right)  \\
&-&  \varphi \sum_{m=even}^{M-2} c_m \tilde{\varphi}_m 
+ \sum_{m=even}^{M-2} \left( \frac{U}{2}	\tilde{\bm{n}}_{m}^2  +   \frac{\omega_m^2}{2U}  \tilde{\varphi}_m^2
\right) 
\eeq
where we have changed the notations, namely $\varphi=\varphi_{-}$ and the conjugate $\bm{n}=\bm{n}_{-}/2$.   
The effective number of bath DOFs is 
\beq
d_{\text{bath}}=\lfloor (M{-}2)/2 \rfloor
\eeq 
Re-writing the bath in the standard Caldeira-Leggett form \Eq{e3} the JCH takes the form
\beq
\mathcal{H} \ \ = \ \
U \bm{n}^2  + \frac{\bar{n}K}{4} \varphi^2 -\bar{n}K^{\prime} \cos\left(\varphi -\Phi\right)
+V_{\text{counter}} + \mathcal{H}_{\text{bath}}  
\eeq
In order to get \Eq{e5} one has to do some algebra with the counter-term:
\beq
V_{\text{counter}} &=& - \varphi^2 \sum_{m=even}^{M-2} \frac{U c_m^2}{2\omega_m^2} 
= -\varphi^2 \frac{\bar{n}K}{2(M-1)} \sum_{m=even}^{M-2} \frac{\sin^2 k_m}{1-\cos k_m} 
\\ 
&=&  -\varphi^2 \frac{\bar{n}K}{M-1} \sum_{m=even}^{M-2} \cos^2 \left(\frac{k_m}{2}\right)
= - \frac{1}{4}\left(\frac{M-3}{M-1}\right) \bar{n}K \, \varphi^2
\eeq

\clearpage

\noindent
{\bf References.-- }


\providecommand{\newblock}{}

\clearpage
\end{document}